\documentclass[useAMS,usenatbib]{mn2e}
\usepackage{graphicx}

\newcommand{\afe}{[\alpha/\mathrm{Fe}]}
\newcommand{\hbeta}{\mathrm{H}\beta} 
\newcommand{\hbetag}{\mathrm{H}\beta_\mathrm{G}} 
\newcommand{\mgb}{Mg\textit{b}} 

\title[The stellar populations of early-type galaxies -- II]{The
  stellar populations of early-type galaxies -- II. The effects of
  environment and mass}
\author[Craig Harrison et al.]{Craig D. Harrison$^{1,2}$\thanks{E-mail:
charrison@ctio.noao.edu}, Matthew Colless$^{3}$, Harald
  Kuntschner$^{4}$, Warrick J. Couch$^{5}$, \newauthor Roberto De
  Propris$^{2}$, Michael B. Pracy$^{5}$\\ 
$^{1}$Research School of Astronomy \& Astrophysics, Australian
  National University, Weston Creek, ACT 2611, Australia\\
$^{2}$Cerro Tololo Inter-American Observatory, Casilla 603, La Serena,
  Chile\\
$^{3}$Anglo-Australian Observatory, PO Box 296, Epping, NSW 2111, Australia\\
$^{4}$Space Telescope European Coordinating Facility, European
  Southern Observatory, Karl-Schwarzschild-Str, 85748 Garching, Germany\\
$^{5}$Centre for Astrophysics \& Supercomputing, Swinburne University
  of Technology, PO Box 218, Hawthorn, VIC 3122, Australia}
\begin{document}

\date{Accepted ... . Received ... ; in original form 2010 January 7}

\pagerange{\pageref{firstpage}--\pageref{lastpage}} \pubyear{2010}

\maketitle

\label{firstpage}

\begin{abstract}
The degree of influence that environment and mass have on the stellar
populations of early-type galaxies is uncertain. In this paper we
present the results of a spectroscopic analysis of the stellar
populations of early-type galaxies aimed at addressing this
question. The sample of galaxies is drawn from four clusters, with
$\left< z \right>=0.04$, and their surrounding structure extending to
$\sim10 R_\mathrm{vir}$. We find that the distributions of the
absorption-line strengths and the stellar population parameters age,
metallicity and $\alpha$-element abundance ratio do not differ
significantly between the clusters and their outskirts, but the tight
correlations found between these quantities and velocity dispersion
within the clusters are weaker in their outskirts. All three stellar
population parameters of cluster galaxies are positively correlated
with velocity dispersion. Galaxies in clusters form a homogeneous
class of objects that have similar distributions of line-strengths and
stellar population parameters, and follow similar scaling relations
regardless of cluster richness or morphology. We estimate the
intrinsic scatter of the Gaussian distribution of metallicities to be
0.3 dex, while that of the $\alpha$-element abundance ratio is 0.07
dex. The $e$-folding time of the exponential distribution of galaxy
ages is estimated to be 900 Myr. The intrinsic scatters of the
metallicity and $\alpha$-element abundance ratio distributions can
almost entirely be accounted for by the correlations with velocity
dispersion and the intrinsic scatter about these relations. This
implies that a galaxies mass plays the major role in determining its
stellar population.
\end{abstract}


\section{Introduction}\label{intro}

The classical model of galaxy formation, the monolithic collapse model
\citep{eggen62, larson74, larson75, tinsley76, arimoto87, bressan94},
proposes that galaxies form in a single massive collapse at high
redshift and that subsequent evolution is purely passive. This model
has been overtaken by the current dominant model of galaxy formation,
the hierarchical merging model \citep{toomre77, searle78, white78},
which proposes that galaxies are built up through mergers: small
objects form first and undergo a series of mergers that build up more
massive objects. A third scenario, the revised monolithic collapse
model \citep[e.g.][]{merlin06}, proposes that galaxies form in a
number of rapid mergers at high redshift before evolving passively.

Galaxy clusters originate from the most extreme density fluctuations,
where galaxy formation and evolution is expected to proceed at an
accelerated rate. Moreover, the stellar populations of galaxies moving
into a cluster environment are expected to be modified via
interactions with other galaxies and the dense intra-cluster
medium. These interactions can be roughly divided into two broad
classes: local processes including mergers \citep{toomre77} and tidal
interactions \citep{mastropietro05}; and global processes including
ram pressure stripping \citep{gunn72}, interactions with the cluster
tidal field \citep{bekki99}, harassment \citep{moore99}, and
strangulation \citep{larson80}. Local processes are more efficient in
galaxy groups, where relative velocities are lower, while global
processes are more efficient in clusters where the frequency of
interaction is higher.

Correlations are therefore expected between galaxy observables and
environment \citep{kauffmann96, kauffmann98}, and have been found for
galaxy morphology \citep{davis76, dressler80, postman84, balogh98},
colour \citep{blanton05}, S\'ersic index \citep{blanton05,
  hashimoto99}, star-formation rate \citep{lewis02, gomez03,
  boselli06}, and spectral type \citep{norberg02}. However, in a
comprehensive study of the dependence of galaxy observables on
environment within the SDSS, \citet{blanton05} suggest that the
structural properties of galaxies are less dependent on environment
than their masses and star-formation histories.

To add to this debate, Fundamental Plane studies have found no
differences between field galaxies more massive than $2\times10^{11}
M_\odot$ and their counterparts in clusters at the same redshift
\citep{treu99, treu01, vandokkum01}, while less massive galaxies were
found to be younger in the field \citep{treu02, vandokkum03,
  vanderwel04, vanderwel05, treu05a, treu05b}. This might imply that
star formation in field galaxies occurs first in the most massive
galaxies then progressively in less massive galaxies, and that mass
rather than environment governs the overall growth.

The Lick system of absorption-line indices \citep{burstein84, faber85,
  burstein86, gorgas93, worthey94a, trager98} and associated models
\citep[e.g.][]{worthey94b, thomas03a, thomas04}, from which the
stellar population parameters (SPPs) age, metallicity ([Z/H]), and
$\alpha$-element abundance ratio ($\afe$) can be estimated, provide an
excellent method with which to study the role of mass and environment
in determining galaxy properties.

The line-strengths of early-type galaxies are found to be correlated
with velocity dispersion ($\sigma$). Lick indices that are more
sensitive to [Z/H] effects, e.g.\ \mgb, are positively correlated
\citep{burstein88, bender93, ziegler97, colless99a, kuntschner00}
while those indices more sensitive to age effects, e.g.\ $\hbeta$, are
negatively correlated \citep{fisher95, fisher96, jorgensen97,
  jorgensen99a, kuntschner00, caldwell03, bernardi03}. Although both
elements are sensitive to [Z/H], the slope of the correlation between
Mg and velocity dispersion is found to be much steeper than that of Fe
\citep{worthey92a, fisher95, greggio97, jorgensen99a, kuntschner00,
  terlevich02}.

These index--$\sigma$ correlations suggest that the SPPs should also
be correlated with velocity dispersion. Given the difference in the
slopes of the Mg--$\sigma$ and Fe--$\sigma$ relations, a correlation
between $\afe$ and velocity dispersion was expected and has been
confirmed \citep{trager00a, proctor02, thomas02, mehlert03,
  thomas05}. If the $\alpha$-element enhancement is due to the
timescale of star formation, and velocity dispersion is a proxy for
mass, then this correlation indicates that more massive galaxies form
their stars on shorter timescales than less massive galaxies.

The [Z/H] of a galaxy is also found to correlate with velocity
dispersion \citep{greggio97,thomas05}, with massive galaxies being
more metal-rich. Such a relation is a natural consequence of galactic
wind models \citep[e.g.][]{arimoto87}, which show that the larger
gravitational potential of massive galaxies allows them to better
retain their heavy elements.

Whether a correlation exists between age and velocity dispersion is
still uncertain. Early studies found no significant correlation
\citep{trager00a, kuntschner01, terlevich02}, but recent studies have
detected a weak but significant correlation having large scatter
\citep{proctor02, proctor04a, proctor04b, thomas05}, with more massive
galaxies being older. This trend, which would indicate that more
massive galaxies formed their stellar content earlier, is in agreement
with the concept of down-sizing \citep{cowie96} and the latest
semi-analytic models of galaxy formation \citep[e.g.,][]{delucia06}.

Early-type galaxies in low-density environments exhibit small
differences to those in clusters: at a given luminosity, the
early-type galaxies in low-density regions are $\sim1-3$ Gyr younger
and $0.1-0.2$ dex more metal-rich than those in clusters
\citep{trager00a, poggianti01, kuntschner02, terlevich02, caldwell03,
  proctor04a, thomas05, sanchez06}. In conflict with the above
results, \citet{gallazzi06}, using a large sample of early-type
galaxies from the SDSS, found evidence that galaxies in low-density
environments were less metal-rich than those in high-density
environments. Intriguingly, there appears to be no environmental
dependence of $\afe$ \citep{kuntschner02, thomas05}, indicating that
star formation in galaxies of a given mass occurs on the same
timescale whether they are located in high-density or low-density
environments.

While it appears that the stellar populations of galaxies change from
low-density environments to high-density environments, the density
threshold at which this change occurs is uncertain. Studies of the
star-formation rate of galaxies in and around clusters \citep{lewis02,
  gomez03} find an increase in the star-formation rate with increasing
distance from the cluster centre, converging to the mean field rate at
distances greater than $\sim3 R_\mathrm{vir}$. The critical projected
density at which suppression of star-formation begins is uncertain but
the environmental influences on galaxy properties are believed not to
be restricted to cluster cores, being effective in all groups where
the density exceeds the critical value. The observed low rates of star
formation well beyond the virialised cluster rule out physical
processes associated with extreme environments (such as ram pressure
stripping of disk gas) being completely responsible for the variations
in galaxy properties with environment.

This paper is the second in a series of papers aimed at studying the
effects of environment and mass on the stellar populations of
early-type galaxies. The first paper described our sample selection,
observations, data reductions, and method of measuring line-strengths
and estimating stellar population parameters within the framework of
the Lick system. In this paper we present the results of this study
utilising data from four clusters and their surrounds. The data
extends to $\sim 10 R_\mathrm{vir}$, allowing us to probe the in-fall
regions of the clusters, which to date have been poorly studied. The
layout of the remainder of the paper is as follows. \S \ref{obs}
briefly describes the sample, observations and reductions, measurement
of Lick indices and SPP estimation. Analysis of the absorption-line
strengths is detailed in \S \ref{indices} and the distributions of the
SPPs in \S \ref{params}. The correlations between the SPPs and
velocity dispersion are discussed in \S \ref{scale} while the
contribution from these correlations to the intrinsic scatter in the
parameter distributions are investigated in \S \ref{scatter}. We
discuss the results of this study in \S \ref{discuss} and present a
summary of the work in \S \ref{summary}.

A Hubble parameter $H_0=70$ km s$^{-1}$ Mpc$^{-1}$, matter density
parameter $\Omega_M=0.3$ and dark energy density parameter
$\Omega_\Lambda=0.7$ are adopted throughout this work.

\section{From Observations to Stellar Population
  Parameters}\label{obs}

Details of the observations, data reductions, measurement of Lick
indices and estimation of the corresponding SPPs are given in Harrison
et al.\ (2010, submitted; hereafter referred to as Paper I). Therefore
only brief descriptions will be given here.

Observations of galaxies from four clusters (A930, A1139, A3558 and
Coma) of varying richness and morphology were made with 2dF during the
nights of 19--21 April 2002. Observations of galaxies from the same
four clusters and the structures around them were made with 6dF during
the nights of 6--8 March 2003. The 6dF sample contained some galaxies
in common with the 2dF sample, but mostly consisted of galaxies in the
outer regions of each cluster. In two of the fields (A1139 and A930),
the observations were offset from the cluster centre, allowing
galaxies with cluster-centric distances of up to $\sim 19$
$h_{70}^{-1}$ Mpc to be studied. The set-up of the two instruments can
be found in Paper I.

A brief discussion of the way galaxies were classified as early types
or not is warranted here. This was done spectroscopically with
galaxies that showed signs of H$\alpha$ (EW$< -3.8$ \AA) or
[OIII]$\lambda$5007\,\AA\ (EW$< -0.4$ \AA) emission being classified
as star forming. We chose to classify a galaxy by its spectrum because
comparisons to SSP models are (almost) meaningless in galaxies with
significant emission. In-fill of the $\hbeta$ absorption feature by
nebular emission results in weaker $\hbeta$ line-strengths and
incorrectly older ages. Some methods used to correct for this nebular
in-fill \citep[e.g.][]{gonzalez93,trager00a,trager00b} have proved
unsatisfactory, while others \citep{sarzi06} can only correct for
relatively weak emission. For these reasons we eliminated all galaxies
with signs of emission from our sample of early types. It must be
noted that, in doing so, it is possible that we exclude from our
stellar population analysis early-type galaxies that have had star
formation triggered by the cluster environment.

Other methods of classifying galaxies are not
problem-free. Classifying by morphology is highly subjective and
samples selected by colour-magnitude cuts still contain contamination
by galaxies that would have been morphologically classified as late
types. More importantly, these methods do not eliminate all
emission-line galaxies, up to 30\% of red sequence galaxies can show
signs of LINER emission \citep{graves07}, and so the interpretation of
a stellar population analysis for these galaxies would be difficult.

However, the absence of emission lines does not guarantee that a
galaxy has not had significant star formation in its recent past and
so it is possible that our sample of early types contains a small
number of post-starburst galaxies. The intrinsic scatters in the line
strength--$\sigma$ (Section \ref{index sigma}) and stellar population
parameter--$\sigma$ relations (Section \ref{low clus rels}) are
comparable to published results implying that if our sample contains
recently star-forming galaxies then it is in no greater number than
previous samples.

The basic reduction steps, such as bias subtraction, spectrum
extraction, flat-fielding, wavelength calibration, fibre throughput
determination and correction, and sky subtraction, were performed with
the purpose-built data reduction packages 2dfdr \citep{colless01} and
6dfdr \citep{jones04}. Redshifts were measured using the program runz
\citep{colless01} while the IRAF task fxcor was used to measure
velocity dispersions.

Line-strengths were measured and transformed to the Lick system
closely following procedures outlined in a number of papers
\citep[e.g.][]{gonzalez93, fisher95, worthey97, trager98,
  kuntschner00}. The spectra were broadened to the Lick resolution
($\sim$ 9 \AA\ FWHM), and the line-strengths were measured using the
program indexf \citep{cardiel98}.

A number of corrections were then applied to the measured
line-strengths to fully calibrate them to the Lick system. These
corrections include a velocity dispersion correction to account for
the change in line-strength caused by velocity broadening, an aperture
correction to account for the different linear sizes subtended by the
different fibres at different redshifts, and, finally, applying any
offsets that may arise due to the fact that the Lick spectra were not
flux calibrated. This resulted in the measurement of the line-strength
indices C$_2$4668, $\hbeta$, $\hbetag$, [OIII]$_1$, [OIII]$_2$,
Fe5015, Mg$_1$, Mg$_2$, \mgb, Fe5270, Fe5335 and Fe5406 of the stellar
populations within the inner $\sim1$ kpc of each galaxy.
 
The age, [Z/H], and $\afe$ of a galaxy were obtained by comparing our
measured line-strengths to the models of \citet{thomas03a} and
accepting the combination of SPPs of the model with the smallest
$\chi^2$ \citep{proctor04b}. We interpolate the models logarithmically
in steps of 0.02 dex in age and [Z/H], and 0.01 dex in $\afe$. Errors
on the parameters were estimated by using the constant $\chi^2$
boundaries as confidence limits \citep[see][]{press92}.

In summary, we measured velocity dispersions, redshifts, and
line-strengths for a total of 416 galaxies: 158 of these are
early-type galaxies from Coma, A1139, A3558, or A930 that are located
within the Abell radius (i.e.\ at a projected radial distance $\le2$
$h_{70}^{-1}$ Mpc) and comprise our cluster sample; 87 are early-type
galaxies from the outskirts of these clusters at projected radial
distances $>2$ $h_{70}^{-1}$ Mpc and comprise the cluster-outskirts
sample; 168 galaxies were deemed to be star-forming, and make up the
emission-line sample; the 3 remaining galaxies lacked the information
necessary to classify them as either an early-type or emission-line
galaxy. Of these 416 galaxies SPPs were estimated for 219; 142 in the
cluster sample and 77 in the cluster-outskirts sample.

\section{Absorption-Line Analysis}\label{indices}

\subsection{\boldmath Line-strength--$\sigma$ relations in cluster early-type
  galaxies}\label{index sigma}
 
\begin{table*}
\caption{The line-strength--$\sigma$ fits, Spearman rank correlation
  statistics, probabilities of a correlation, and the intrinsic
  scatter ($\delta_\mathrm{intr}$; in mag) for the cluster galaxies.}
\label{index sigma corr}
\begin{tabular}{lr@{}lr@{}lr@{.}lcc}
\hline
Index & \multicolumn{2}{c}{Zero-point} & \multicolumn{2}{c}{Slope} & 
\multicolumn{2}{c}{$r_\mathrm{S}$} & Probability & $\delta_\mathrm{intr}$\\
\hline
$\mathrm{C}_24668^\prime$ & $-$ & $0.053 \pm
0.009$ &  & $0.067 \pm 0.004$ &     0 & 546 & 1.000 & $0.0096 \pm 0.0011$\\ 
$\hbeta^\prime$ &  & $0.167 \pm 0.009$ & $-$ & $0.044 \pm 0.004$ & 
$-0$ & 479 & 1.000 & $0.0075 \pm 0.0012$\\
$\hbetag^\prime$ &  & $0.238 \pm 0.011$ & $-$ & $0.054 \pm 0.005$ &
$-0$ & 442 & 1.000 & $0.0120 \pm 0.0022$\\
$\mathrm{Fe}5015^\prime$ &  & $0.046 \pm 0.008$ &  & $0.013 \pm 0.003$
&     0 & 284 & 0.999 & $0.0035 \pm 0.0008$\\ 
$\mathrm{Mg}_1$ & $-$ & $0.285 \pm 0.018$ &  & $0.180 \pm 0.008$ 
&     0 & 626 & 1.000 & $0.0198 \pm 0.0021$\\
$\mathrm{Mg}_2$ & $-$ & $0.402 \pm 0.032$ &  & $0.299 \pm 0.014$ 
&     0 & 662 & 1.000 & $0.0200 \pm 0.0001$\\
$\mathrm{\mgb}^\prime$ & $-$ & $0.130 \pm 0.012$ &  & $0.129 \pm 0.005$ 
&     0 & 675 & 1.000 & $0.0094 \pm 0.0013$\\
$\mathrm{Fe}5270^\prime$ &  & $0.044 \pm 0.008$ &  & $0.016 \pm 0.004$ 
&     0 & 342 & 0.999 & $0.0045 \pm 0.0012$\\
$\mathrm{Fe}5335^\prime$ &  & $0.008 \pm 0.011$ &  & $0.028 \pm 0.005$ 
&     0 & 352 & 1.000 & $0.0081 \pm  0.0009$\\
$\mathrm{Fe}5406^\prime$ &  & $0.043 \pm 0.011$ &  & $0.010 \pm 0.005$ 
&     0 & 166 & 0.921 & $0.0035 \pm 0.0012$\\
\hline
\end{tabular}
\end{table*}

In early-type galaxies the strengths of certain absorption features
are observed to be correlated with velocity dispersion. Such
correlations are important as they provide links between a galaxy's
dynamical and chemical evolution; i.e.\ between galaxy mass,
metallicity, and abundance ratios.

Historically, the Mg--$\sigma$ relation was studied using Mg$_2$,
which is measured in magnitudes of absorbed flux. More recent studies
\citep[e.g.][]{colless99a, kuntschner01} have used \mgb, which is a
more reliable index, as it is narrower, but is measured as an
equivalent width in \AA. These studies converted indices measured in
equivalent widths to magnitudes \citep[for the conversion
  see][]{colless99a}, primarily to allow comparison with older studies
that used Mg$_2$. We continue this practice and indices that have been
converted to magnitudes will be denoted with a prime symbol
(e.g.\ \mgb$^\prime$).

The variations of the index line-strengths with velocity dispersion
for the cluster sample are shown in Figure \ref{index sigma
  rels}. These plots contain the combined data from all four clusters,
as Kolmogorov-Smirnov (KS) 2D two-sample tests indicate consistency of
the line-strength--$\sigma$ distributions between clusters, in that
most distributions differed only at the 1--2$\sigma$ level. That these
relations are found to be consistent in the four clusters, which span
the ranges of Abell richness classes and Bautz-Morgan (BM)
morphologies, is remarkable. This suggests that the cluster
environment has little effect on the line-strengths of early-type
galaxies (see \citet{colless99a} for a study of the Mg--$\sigma$
relation as a function of cluster environment). Table \ref{index sigma
  corr} shows the results of linear fits to the data, accounting for
errors in both quantities, along with the Spearman rank correlation
statistic ($r_\mathrm{S}$), the probability that the two quantities
are correlated and the estimated intrinsic scatter about each
relation.

We assume the intrinsic scatter about these relations is Gaussian and
estimate it by maximising the logarithmic likelihood 
\begin{equation}\label{log like}
\mathcal{\log L}=\sum_{i=1}^n \log
  \left[\frac{1}{\sqrt{2\pi\delta_i^2}} \exp \left(\frac{-\Delta
  I_i^2}{2\delta_i^2}\right)\right]\hspace{1ex},
\end{equation}
where the $\Delta I_i$ are the residuals to the fit for index $I$ and
$\delta_i^2=\delta I^2_i+(a_I\,\delta\log
\sigma_i)^2+\delta^2_\mathrm{intr}$, where $\delta I_i$ and $\delta
\log \sigma_i$ are the individual errors on the index $I$ and
$\log\,\sigma$, $a_I$ is the slope of the relation for index $I$, and
$\delta_\mathrm{intr}$ is the estimated intrinsic scatter. These
estimated intrinsic scatters are listed in Table \ref{index sigma
  corr}.

With the exception of Fe5406, correlations are found at $> 3\sigma$
confidence level for all indices. Strong correlations ($r_\mathrm{S} >
0.5$) are found for C$_24668$, Mg$_1$, Mg$_2$ and \mgb, while a weak
correlation ($r_\mathrm{S}< 0.3$) is found for Fe5015. Except for
$\hbeta$ and $\hbetag$, which are moderately anti-correlated, all
other indices are moderately correlated. The slope of the fit to the
Fe5406 data is very flat and consistent with being zero. $\hbeta$ and
$\hbetag$ are the only indices used here that are more sensitive to
age effects and the only ones that exhibit an anti-correlation; all
the other indices are more sensitive to [Z/H] effects and either
exhibit a positive correlation or no correlation.

Young stellar populations exhibit strong Balmer absorption features,
which weaken with age. Therefore, a simple interpretation of the
anti-correlation of $\hbeta$ and $\hbetag$ with velocity dispersion is
that more massive galaxies are older. The simple interpretation of the
positive correlation between the metal-sensitive indices and velocity
dispersion is that more massive galaxies are more metal-rich. The
usual explanation of this trend is that massive galaxies are better
able to retain their heavy elements due to their larger gravitational
potential, a scenario that arises naturally in galactic wind models
\citep[e.g.][]{arimoto87}. Alternatively, a variable IMF could lead to
a similar mass-[Z/H] relation. \citet{koeppen07} have developed a
model where the effective upper mass limit of stars is lower in
galaxies with a low star-formation rate. This reduces the number of
SNII, and hence the [Z/H], in low mass galaxies and leads to a similar
mass-[Z/H] relation.

\begin{figure*}
\includegraphics[width=1.0\textwidth]{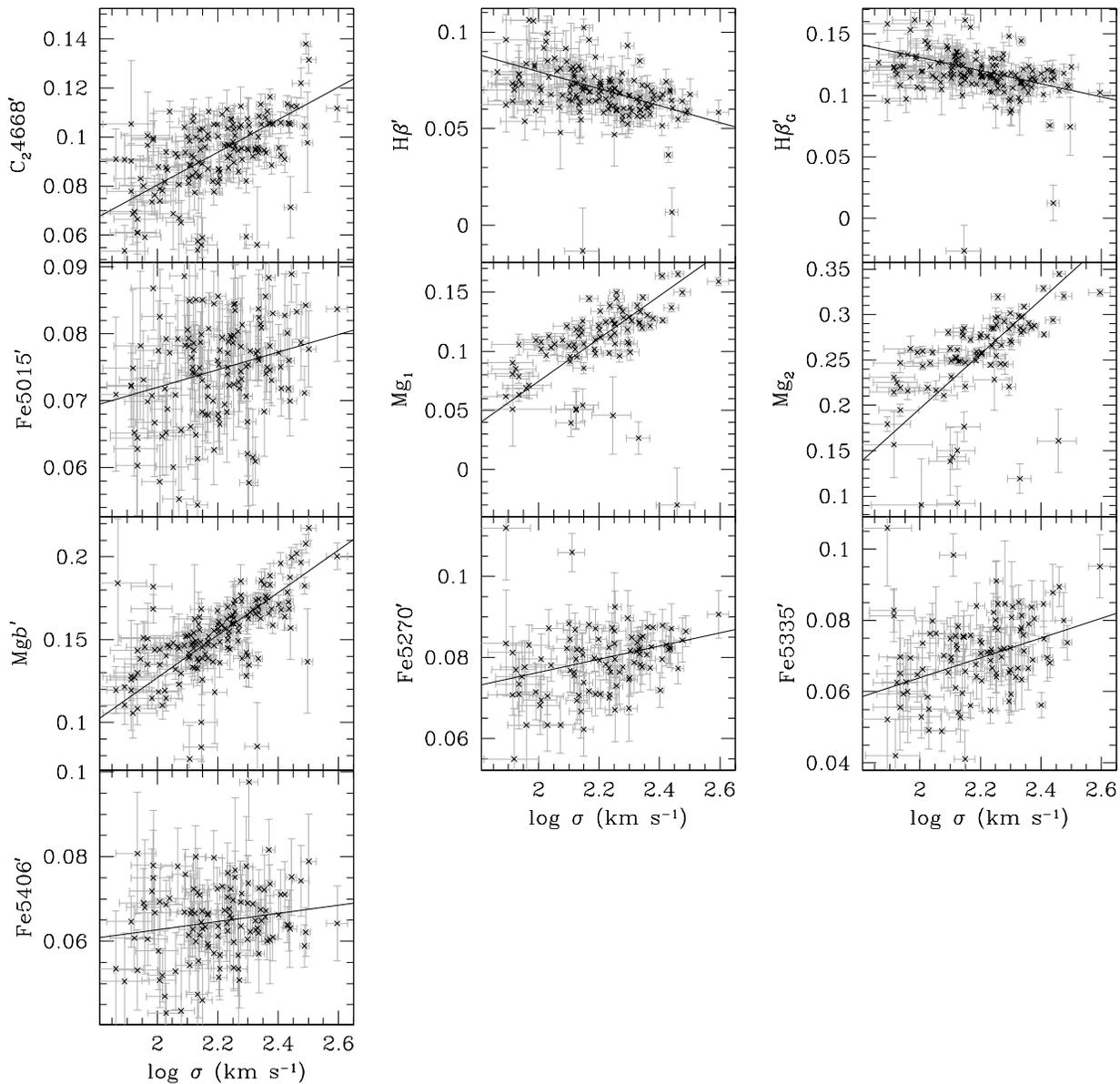}
\caption{The variations (in magnitudes) of the index line-strengths
  with velocity dispersion for the cluster sample. Most indices
  exhibit a tight relation with strong correlations found for
  C$_2$4668, Mg$_1$, Mg$_2$ and \mgb, while $\hbeta$ and $\hbetag$
  exhibit moderate anti-correlations.}
\label{index sigma rels}
\end{figure*}

\begin{table*}
\caption{A comparison of the line-strength--$\sigma$ relations to
  those found in the literature.}
\label{index sigma comp}
\begin{tabular}{lr@{.}lr@{.}lr@{.}lr@{.}lr@{.}lr@{.}l}
\hline
Reference            &  \multicolumn{4}{c}{C$_2$4668$^\prime$}     & \multicolumn{4}{c}{$\hbeta$$^\prime$}       & \multicolumn{4}{c}{Fe5015$^\prime$}\\
                     &  \multicolumn{2}{c}{Zero-point}             &  \multicolumn{2}{c}{Slope}                  &  \multicolumn{2}{c}{Zero-point}
                     &  \multicolumn{2}{c}{Slope}                  &  \multicolumn{2}{c}{Zero-point}             &  \multicolumn{2}{c}{Slope}\\
\hline
This study           & $-0$ & 053$\pm$0.009 & 0    & 067$\pm$0.004 & 0    & 167$\pm$0.009 & $-0$ & 044$\pm$0.004 & 0 & 046$\pm$0.008    & 0    & 013$\pm$0.003\\
\citet{kuntschner00} & $-0$ & 110$\pm$0.042 & 0    & 090$\pm$0.018 & 0    & 106$\pm$0.015 & $-0$ & 020$\pm$0.007 & 0 & 002$\pm$0.019    & 0    & 036$\pm$0.008\\
\citet{nelan05}      & \multicolumn{2}{c}{} & 0    & 075$\pm$0.002 & \multicolumn{2}{c}{} & $-0$ & 041$\pm$0.001 & \multicolumn{2}{c}{} & 0    & 015$\pm$0.001\\
\citet{clemens06}    & 0    & 044$\pm$0.003 & 0    & 025$\pm$0.003 & 0    & 227$\pm$0.002 & $-0$ & 071$\pm$0.002 & 0 & 127$\pm$0.002    & $-0$ & 023$\pm$0.002\\
\citet{sanchez06}    & $-0$ & 071$\pm$0.018 & 0    & 067$\pm$0.008 & 0    & 087$\pm$0.009 & $-0$ & 012$\pm$0.004 & 0 & 025$\pm$0.011    & 0    & 021$\pm$0.005\\
\citet{ogando08}     & \multicolumn{4}{c}{}                        & 0    & 156$\pm$0.013 & $-0$ & 042$\pm$0.006 & 0 & 053$\pm$0.010	& 0    & 011$\pm$0.004\\
\citet{matkovic09}   & $-0$ & 044$\pm$0.014 & 0    & 060$\pm$0.014 & 0    & 116$\pm$0.009 & $-0$ & 024$\pm$0.009 & 0 & 060$\pm$0.007    & 0    & 005$\pm$0.007\\
\hline
Reference            & \multicolumn{4}{c}{Mg$_1$}                  &  \multicolumn{4}{c}{Mg$_2$}                 &  \multicolumn{4}{c}{\mgb$^\prime$}\\
                     &  \multicolumn{2}{c}{Zero-point}             &  \multicolumn{2}{c}{Slope}                  &  \multicolumn{2}{c}{Zero-point}
                     &  \multicolumn{2}{c}{Slope}                  &  \multicolumn{2}{c}{Zero-point}             &  \multicolumn{2}{c}{Slope}\\
\hline
This study           & $-0$ & 285$\pm$0.018 & 0    & 180$\pm$0.008 & $-0$ & 402$\pm$0.032 & 0    & 299$\pm$0.014 & $-0$ & 130$\pm$0.012 & 0    & 129$\pm$0.005\\
\citet{kuntschner00} & $-0$ & 158$\pm$0.035 & 0    & 136$\pm$0.015 & $-0$ & 127$\pm$0.054 & 0    & 191$\pm$0.023 & $-0$ & 056$\pm$0.044 & 0    & 102$\pm$0.020\\
\citet{nelan05}      & \multicolumn{2}{c}{} & 0    & 121$\pm$0.003 & \multicolumn{2}{c}{} & 0    & 189$\pm$0.003 & \multicolumn{2}{c}{} & 0    & 134$\pm$0.002\\
\citet{clemens06}    & $-0$ & 245$\pm$0.010 & 0    & 163$\pm$0.004 & $-0$ & 198$\pm$0.013 & 0    & 208$\pm$0.006 & $-0$ & 155$\pm$0.007 & 0    & 142$\pm$0.007\\
\citet{sanchez06}    & \multicolumn{8}{c}{}                                                                      & $-0$ & 050$\pm$0.019 & 0    & 091$\pm$0.008\\
\citet{ogando08}     & $-0$ & 159$\pm$0.020 & 0    & 131$\pm$0.009 & $-0$ & 153$\pm$0.029 & 0    & 194$\pm$0.013 & $-0$ & 095$\pm$0.018	& 0    & 112$\pm$0.008\\
\citet{matkovic09}   & $-0$ & 204$\pm$0.036 & 0    & 141$\pm$0.017 & $-0$ & 143$\pm$0.045 & 0    & 178$\pm$0.020 & $-0$ & 023$\pm$0.015 & 0    & 080$\pm$0.015\\
\hline
Reference            &  \multicolumn{4}{c}{Fe5270$^\prime$}        &  \multicolumn{4}{c}{Fe5335$^\prime$}        &  \multicolumn{4}{c}{Fe5406$^\prime$}\\
                     &  \multicolumn{2}{c}{Zero-point}             &  \multicolumn{2}{c}{Slope}                  &  \multicolumn{2}{c}{Zero-point}
                     &  \multicolumn{2}{c}{Slope}                  &  \multicolumn{2}{c}{Zero-point}             &  \multicolumn{2}{c}{Slope}\\
\hline
This study           & 0    & 044$\pm$0.008 & 0    & 016$\pm$0.004 & 0    & 008$\pm$0.011 & 0    & 028$\pm$0.005 & 0    & 043$\pm$0.011 & 0    & 010$\pm$0.005\\
\citet{kuntschner00} & 0    & 024$\pm$0.020 & 0    & 029$\pm$0.009 & $-0$ & 017$\pm$0.020 & 0    & 043$\pm$0.009 & 0    & 023$\pm$0.026 & 0    & 023$\pm$0.012\\
\citet{nelan05}      & \multicolumn{2}{c}{} & 0    & 017$\pm$0.001 & \multicolumn{2}{c}{} & 0    & 023$\pm$0.001 & \multicolumn{2}{c}{} & 0    & 018$\pm$0.001\\
\citet{clemens06}    & 0    & 085$\pm$0.003 & $-0$ & 001$\pm$0.003 & 0    & 014$\pm$0.003 & 0    & 027$\pm$0.003 & $-0$ & 002$\pm$0.003 & 0    & 031$\pm$0.003\\
\citet{sanchez06}    & 0    & 027$\pm$0.007 & 0    & 024$\pm$0.003 & $-0$ & 001$\pm$0.013 & 0    & 035$\pm$0.006 & \multicolumn{4}{c}{}                       \\
\citet{ogando08}     & 0    & 049$\pm$0.008 & 0    & 015$\pm$0.004 & 0    & 029$\pm$0.009 & 0    & 020$\pm$0.004 & 0    & 034$\pm$0.009	& 0    & 018$\pm$0.004\\
\citet{matkovic09}   & 0    & 092$\pm$0.009 & $-0$ & 007$\pm$0.009 & 0    & 069$\pm$0.008 & 0    & 001$\pm$0.008 & \multicolumn{4}{c}{}                       \\ 
\hline
\end{tabular}
\end{table*}

We compare our line-strength--$\sigma$ relations to various estimates
from the literature in Table \ref{index sigma comp}. These estimates
are based on a sample of early-type galaxies from the Fornax cluster
\citep{kuntschner00}, a sample of red-sequence galaxies from numerous
low-$z$ clusters \citep{nelan05}, a magnitude-limited sample of
early-type galaxies drawn from the SDSS \citep{clemens06}, a sample of
early-type galaxies in high-density environments \citep{sanchez06}, a
magnitude-limited sample of early-type galaxies in high-density
environments \citep{ogando08}, and a sample of early-type galaxies
from the core of the Coma cluster \citep{matkovic09}.

Generally, our results compare well with those in the literature. We
find that in the majority of cases the slopes agree at the $2\sigma$
level or better. Of those comparisons that disagree by more than
$3\sigma$, we note that almost half are comparisons to the estimates
of \citet{clemens06} and the rest consist almost entirely of
comparisons of Mg$_1$ and Mg$_2$.

The discrepancy with \citet{clemens06} possibly arises due to
differences in the aperture corrections applied to each dataset. We
correct our line-strength measurements to fixed physical size of
$\sim1$ kpc, independent of the velocity dispersion of the galaxy. The
correction applied by \citeauthor{clemens06}, however, is to a
physical size of $r_\mathrm{e}$/10 and is a function of velocity
dispersion. Confusing the issue is the fact that some of their slopes
are steeper than ours (e.g.\ Fe5406), some are shallower
(e.g.\ C$_2$4668) and some are in the opposite sense (e.g.\ Fe5015).

The discrepancy between our estimated slopes of Mg$_1$ and Mg$_2$ and
those in the literature is a little easier to understand. Both these
indices have very broad definitions and so their line-strengths are
highly-sensitive to the shape of the continuum. The fact that we
divided out our continuum is probably the reason for this
discrepancy. This is one of the reasons why we excluded these two
indices from our fits to the models in estimating the SPPs (see
Section 6 in Paper I).

Most comparisons with \citet{matkovic09} agree at better than the
$3\sigma$ level, the exceptions being Mg$_2$ and \mgb, but since both
samples include early-types from the core of the Coma cluster it is
worthwhile to make a direct comparison of these galaxies
alone. Overall, this results in better agreement between the slopes,
although the change is almost negligible in most cases. Three indices
(C$_2$4668, Fe5270 and Fe5335) show worse agreement but the largest
change in significance is only 0.24$\sigma$. All other indices show
better agreement; again, most changes are small except in the case of
$\hbeta$ where the significance changes by 1.38$\sigma$ from
2.03$\sigma$ to 0.65$\sigma$.

The intrinsic scatters found for our line-strength--$\sigma$ relations
are given in Table \ref{index sigma corr}. Most of our estimates agree
well with those of \citet{kuntschner00} and \citet{sanchez06}, with
perhaps the exceptions of $\hbeta$, Fe5015, Fe5270 and Fe5335. We find
an intrinsic scatter in $\hbeta$ of 0.008 mag (as does
\citeauthor{sanchez06}), which is double that of
\citeauthor{kuntschner00}. For the three Fe indices, our scatters are
larger than those found by \citeauthor{sanchez06} who find
negligible-to-no scatter in these relations. Our estimates for these
indices are in good agreement with that of \citeauthor{kuntschner00},
except for Fe5335 for which we find almost double the scatter of that
found by this author (0.008 mag compared to 0.005 mag).

\begin{figure*}
\includegraphics[width=1.0\textwidth]{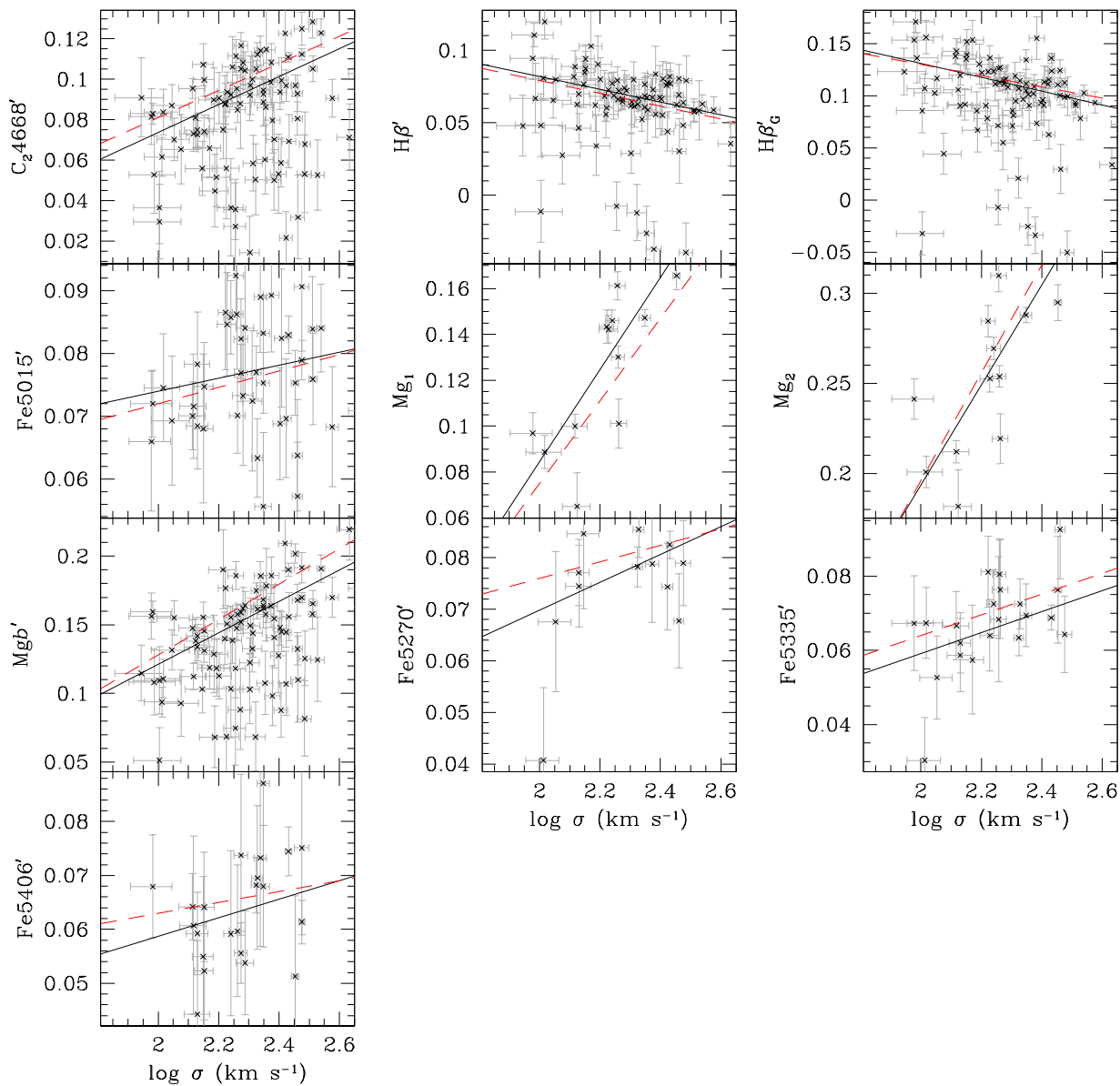}
\caption{The variations (in magnitudes) of the index line-strengths
  with velocity dispersion for the cluster-outskirts sample. The solid
  line is the linear fit to the data while the dashed red line is the
  fit to the cluster sample. The slopes of all relations are
  consistent with those found in the cluster sample.}
\label{index sigma rels env}
\end{figure*}

Given that the galaxies from this study are drawn from four different
clusters, with varying richness classes and BM classifications, the
small intrinsic scatters about the line-strength--$\sigma$ relations
is truly remarkable, and implies that the stellar populations in
cluster early-type galaxies are homogeneous. This result agrees with
that of \citet{colless99a}, who found no correlation between the
Mg--$\sigma$ zero-point and cluster velocity dispersion, X-ray
luminosity, or X-ray temperature \citep[see also][]{bernardi98,
  worthey03, sanchez06}. For galaxies in clusters, \citet{jorgensen96}
found that the Mg--$\sigma$ residuals are weakly correlated with local
density. However, since the residuals are correlated with cluster
velocity dispersion and not projected cluster-centric distance, the
interpretation is of a correlation between the Mg--$\sigma$ zero-point
and cluster mass, albeit a weak one.

\subsection{\boldmath Line-strengths--$\sigma$ relations in the
  cluster-outskirts}

Using the wide field of 6dF and, in the cases of A1139 and A930,
offsetting the field centres with respect to the cluster centres
allowed us to obtain data on galaxies well outside the clusters. We
obtained line-strengths of galaxies out to a projected cluster-centric
radius of $\sim 19$ $h_{70}^{-1}$ Mpc, which corresponds to
$\sim10 R_\mathrm{vir}$. We note that the average distance between
Abell clusters is $\sim 30$ $h_{70}^{-1}$ Mpc.

\begin{table*}
\caption{The line-strength--$\sigma$ fits, Spearman rank correlation
  statistics ($r_\mathrm{S}$), probabilities of a correlation, and
  intrinsic scatters ($\delta_\mathrm{intr}$; in mag) for galaxies in
  the cluster-outskirts sample. The number of standard deviations by
  which the slopes of the cluster-outskirts sample relations differ
  from the cluster sample relations ($\sigma_\mathrm{diff}$) is given
  in the last column.}
\begin{center}
\begin{tabular}{lr@{}lr@{}lr@{.}lccc}
\hline\hline
Index & \multicolumn{2}{c}{Zero-point} &
\multicolumn{2}{c}{Slope} &\multicolumn{2}{c}{$r_\mathrm{S}$} &
Probability & $\delta_\mathrm{intr}$ & $\sigma_\mathrm{diff}$\\
\hline
$\mathrm{C}_24668^\prime$ & $-$ & $0.064 \pm 0.017$ &  & $0.069 \pm
0.007$ &     0 & 260 & 0.985 & $0.0200 \pm 0.0020$ & 0.3\\
$\hbeta^\prime$ &     & $0.171 \pm 0.017$ & $-$ & $0.044 \pm 0.007$ &
$-0$ & 275 & 0.988 & $0.0184 \pm 0.0045$ & 0.0\\
$\hbetag^\prime$ &  & $0.264 \pm 0.021$ & $-$ & $0.066 \pm 0.009$ &
$-0$ & 322 & 0.997 & $0.0200 \pm 0.0003$ & 1.2\\
$\mathrm{Fe}5015^\prime$ &  & $0.053 \pm 0.016$ &  & $0.010 \pm 0.007$
&     0 & 086 & 0.413 & $0.0029 \pm 0.0018$ & 0.4\\
$\mathrm{Mg}_1$ & $-$ & $0.316 \pm 0.056$ &     & $0.200 \pm 0.024$
&     0 & 748 & 0.995 & $0.0142 \pm 0.0038$ & 0.8\\
$\mathrm{Mg}_2$ & $-$ & $0.363 \pm 0.078$ &     & $0.278 \pm 0.034$
&     0 & 650 & 0.978 & $0.0200 \pm 0.0036$ & 0.6\\
$\mathrm{\mgb}^\prime$ & $-$ & $0.108 \pm 0.023$ &     & $0.115 \pm 0.010$
&     0 & 392 & 1.000 & $0.0200 \pm 0.0012$ & 1.3\\
$\mathrm{Fe}5270^\prime$ &  & $0.016 \pm 0.035$ &  & $0.027 \pm 0.015$
&     0 & 410 & 0.814 & $0.0000 \pm 0.0011$ & 0.7\\
$\mathrm{Fe}5335^\prime$ &  & $0.003 \pm 0.030$ &  & $0.028 \pm 0.013$
&     0 & 548 & 0.990 & $0.0000 \pm  0.0020$ & 0.0\\
$\mathrm{Fe}5406^\prime$ &  & $0.024 \pm 0.030$ &  & $0.017 \pm 0.013$
&     0 & 405 & 0.939 & $0.0027 \pm 0.0017$ & 0.5\\
\hline
\end{tabular}
\label{index sigma corr env}
\end{center}
\end{table*}

The trends with velocity dispersion of the line-strengths in the
cluster-outskirts sample (i.e.\ those galaxies outside
$R_\mathrm{Abell}$) are shown in Figure \ref{index sigma rels
  env}. The solid line in each panel is our linear fit to the data
(accounting for errors in both quantities) while the dashed line is
the fit to the cluster galaxies. In Table \ref{index sigma corr env}
we list the details of the fits, the Spearman rank correlation
statistics, the probabilities of a correlation and the estimated
intrinsic scatters about the relations. The last column in this table
shows the number of standard deviations by which the slopes of the
cluster-outskirts sample relations differ from the cluster sample
relations; all indices are found to have slopes consistent with those
found for galaxies in the cluster sample. However, the only indices
that show correlations significant at the $3\sigma$ level in the
cluster-outskirts sample are \mgb\ and $\hbetag$, and while the
strength of the correlation remains the same for $\hbetag$ it is
reduced for \mgb\ from a strong correlation in the cluster sample to a
moderate correlation here.

The change in the Fe indices is interesting. Fe5015 changes from being
weakly correlated in the cluster sample to uncorrelated in the
cluster-outskirts sample. The other three Fe indices remain
moderately-to-weakly correlated (although with reduced significance)
but their slopes are all consistent with being zero. In the cluster
sample only Fe5406 had a slope that was consistent with being zero. In
addition all four Fe indices have decreased intrinsic scatters about
their fits, as does Mg$_1$. C$_24668$, $\hbeta$, $\hbetag$ and
\mgb\ all show an increase in the intrinsic scatter, while for Mg$_2$
it remains the same.

There appears to be a large number of galaxies that have low index
strengths compared to those expected from the relations found in the
cluster sample. For Fe5015 the spread is consistent with what is found
in the cluster sample. The galaxies with low $\hbeta$ and $\hbetag$
strengths, which also are found in the cluster sample but in smaller
numbers, are possibly those that suffer from nebular in-fill. Most of
the galaxies in the cluster-outskirts sample were observed with 6dF
and so have no details of H$\alpha$ emission because the wavelength
coverage was insufficient. Therefore, it is not unexpected that the
level of this possible contamination by star-forming galaxies in this
sample is slightly higher than in the cluster sample.

In the cases of $\mathrm{C}_24668^\prime$ and \mgb, there does appear
to be a real sub-population of galaxies that are genuinely offset from
the relations found in the cluster sample, having weaker
line-strengths for a given velocity dispersion. This is possible if
galaxies in clusters find it easier to retain their heavier elements
than those in their outskirts, due to the dense intra-cluster medium
retarding the development of galactic winds and the expulsion of the
heavier elements.

Another possibility is that the weakened line-strengths are caused by
the aperture correction. Early type galaxies exhibit strong
metallicity and age gradients \citep[e.g.][]{davies93} and indices
measured in massive galaxies will sample a smaller fraction of the
effective radius than those in less massive galaxies. If lower mass
galaxies are found preferentially in the cluster outskirts then this
could lead to a change in the relations in the two regions. These
aperture corrections are necessary to allow a fair comparison between
galaxies in clusters at different redshifts and between galaxies in
the same cluster but observed with fibres that subtend different
angles (2dF$\sim 2.1\arcsec$ and 6dF$\sim 6.7\arcsec$). However, we do
not find that lower mass galaxies are preferentially found in the
cluster-outskirts sample. In fact, while the distributions of
magnitudes are similar in shape in both regions (because we
deliberately targeted brighter galaxies), the peak of the distribution
in the core is shifted to fainter magnitudes relative to the outskirts
sample by $\sim1$\,mag, i.e.\ on average the cluster-outskirts
galaxies are brighter. Therefore this can not be the reason for the
weakening of the relations.

\subsection{\boldmath The radial distribution of star-forming galaxies
  in clusters}

In the analyses so far, the included galaxies were limited to those
that showed no significant sign of on-going star formation, as
determined from H$\alpha$ or [OIII]$\lambda$5007\AA\ emissions. This
allows us to be reasonably confident that their $\hbeta$ absorption
feature is free from in-filling caused by nebular emissions. However,
from these previously excluded galaxies we can determine where in the
cluster and its surrounds star-forming galaxies reside.

\begin{figure*}
\includegraphics[width=1.0\textwidth]{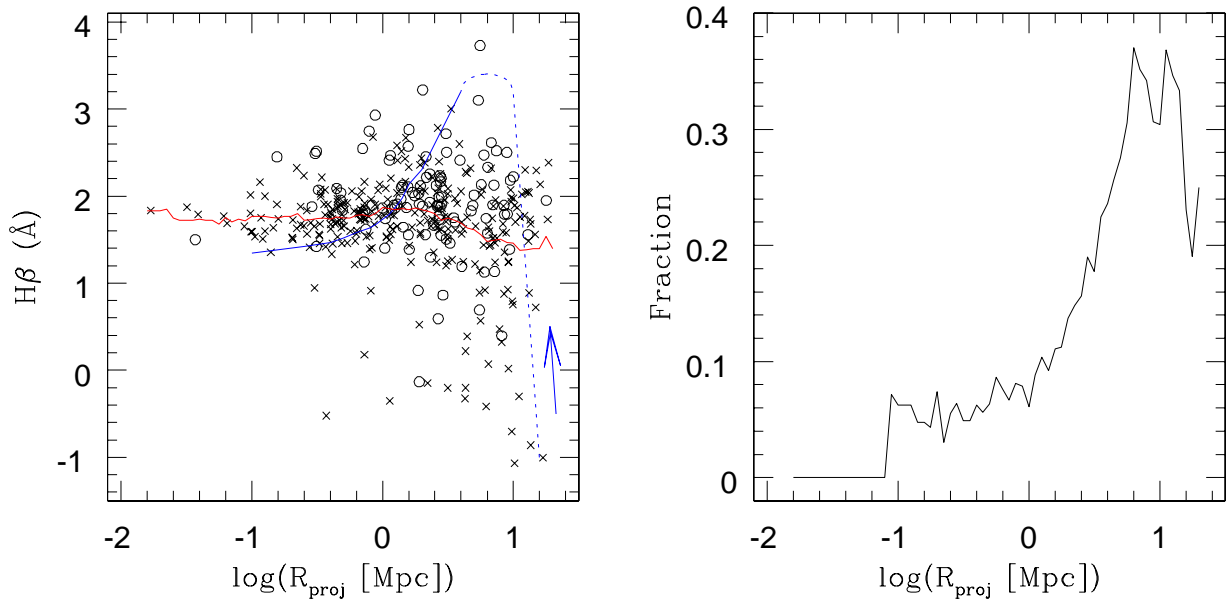}
\caption{Left: The variation in $\hbeta$ line-strength as a function
  of projected cluster-centric radius. Circles represent early-type
  galaxies and crosses represent galaxies that are classified as
  emission-line galaxies. The red line shows how the mean $\hbeta$
  line-strength changes with radius and indicates that on average
  galaxies in the outskirts of clusters are $\sim 0.4$ \AA\ weaker in
  $\hbeta$ than those in the centre of clusters. The blue line shows
  how the $\hbeta$ line-strength evolves in a galaxy as it ages after
  an instantaneous burst of star formation (the direction is given by
  the blue arrow; see text for details). Right: The fraction of
  galaxies with $\hbeta \le 1.4$ \AA\ as a function of projected
  cluster-centric radius.}
\label{hb vs radius}
\end{figure*}

In the left panel of Figure \ref{hb vs radius} we show, for our entire
sample of galaxies, the $\hbeta$ line-strengths as a function of
projected cluster-centric distance. Circles represent galaxies that
are classified as early-types, while crosses represent emission-line
galaxies (and were up to this point excluded from our analysis). The
distribution of $\hbeta$ line-strengths forms a ridge at $\sim 1.8$
\AA. Most of the galaxies with $\hbeta$ line-strengths weaker than
$\sim 1.4$ \AA\ are those classified as emission-line galaxies,
although a few galaxies classified as early-types also lie below this
ridge line. It is possible that these galaxies have weak $\hbeta$ due
to in-filling and yet show no significant sign of
[OIII]$\lambda$5007\AA\ emission. This confirms that while correcting
for $\hbeta$ emission by the strength of the
[OIII]$\lambda$5007\AA\ emission line \citep{gonzalez93, trager00a,
  trager00b} may be acceptable in a statistical sense, it is not
reliably applicable to individual galaxies, which may show moderate to
strong $\hbeta$ emission and little or no
[OIII]$\lambda$5007\AA\ emission \citep{nelan05}.

Figure \ref{hb vs radius} shows an increasing scatter in $\hbeta$
line-strengths with increasing radius. Not only is there a scatter to
lower values (an indication of older ages or possibly $\hbeta$
in-filling in galaxies that have experienced recent star formation),
there is also a scatter to higher values, and thus younger ages, for
early-type galaxies. The red line in the left panel of Figure \ref{hb
  vs radius} shows how the mean H$\beta$ line-strength (calculated for
galaxies within a 0.4 dex radius bin at steps of 0.05 dex) changes
with radius. We see that within a radius of $\sim 2$ $h_{70}^{-1}$ Mpc
(i.e.\ $\sim R_\mathrm{Abell}$) the mean value is a constant ($\sim
1.8$ \AA), while outside this radius it steadily decreases to a value
of $\sim 1.4$ \AA. Since the oldest isochrone (15 Gyr) in the
\citet{thomas03a}\ models has $\hbeta$ strengths $\sim 1.6$--2 \AA\ we
conclude that these low line-strengths, for the galaxies classified as
early types, represent nebular in-fill in recently star-forming
galaxies, which are preferentially found in the outer regions of the
cluster environment. This conclusion is consistent with a previous
study of galaxy star-formation rate and environment by
\citet{lewis02}, which found increasing star-formation rates with
increasing distance from the cluster centre that converge to field
rates at distances greater than $\sim 3 R_\mathrm{vir}$ \citep[see
  also][]{gomez03,kauffmann04}.
 
The distribution of galaxies in this figure can be thought of as
tracking the evolution of a star-forming galaxy. A galaxy falling into
the cluster undergoes a burst of star formation, which shifts the
galaxy to $\hbeta$ emission (i.e.\ negative line-strengths). As the
episode of star formation progresses and the galaxy moves deeper into
the cluster, the galaxy gradually shifts to stronger $\hbeta$
absorption (positive line-strengths). Then, once star formation has
ceased, the galaxy settles back onto the ridge. This evolution is
represented schematically by the blue line in the figure. The dotted
segment of the line represents the galaxy as it moves from being a
star-forming galaxy to a post-starburst galaxy (the $\hbeta$ emission
line-strengths are indicative only). The solid segment of the line
shows how, according to the \citeauthor{thomas03a}\ models, the
$\hbeta$ line-strength changes in a galaxy with [Z/H]=0.35 dex and
$\afe$=0.2 dex as it ages from 1 Gyr to 15 Gyr. The mapping between
projected cluster-centric radius and time is arbitrary. This figure is
similar to Figure 10 in \citet{couch87}, which shows the evolutionary
track of a star-bursting galaxy in H$\delta$--colour space.

The right panel of Figure \ref{hb vs radius} shows the fraction of
galaxies (in the same bins as used for calculating the mean in the
left panel) that have line-strengths less than 1.4 \AA. In the core of
the cluster none of the galaxies fall below this value, while at large
radii the fraction increases rapidly to the limit of our data, where
$\sim 40\%$ of galaxies have line-strengths less than this value. We
conclude that the cluster core is relatively free from young galaxies
and galaxies that have experienced recent star formation, and that
these galaxies are found more commonly outside $R_\mathrm{Abell}$.

\section{Stellar Population Parameter Distributions}\label{params}

\subsection{\boldmath The cluster distributions}
\label{low clus dist}

\begin{figure*}
\includegraphics[width=1.0\textwidth]{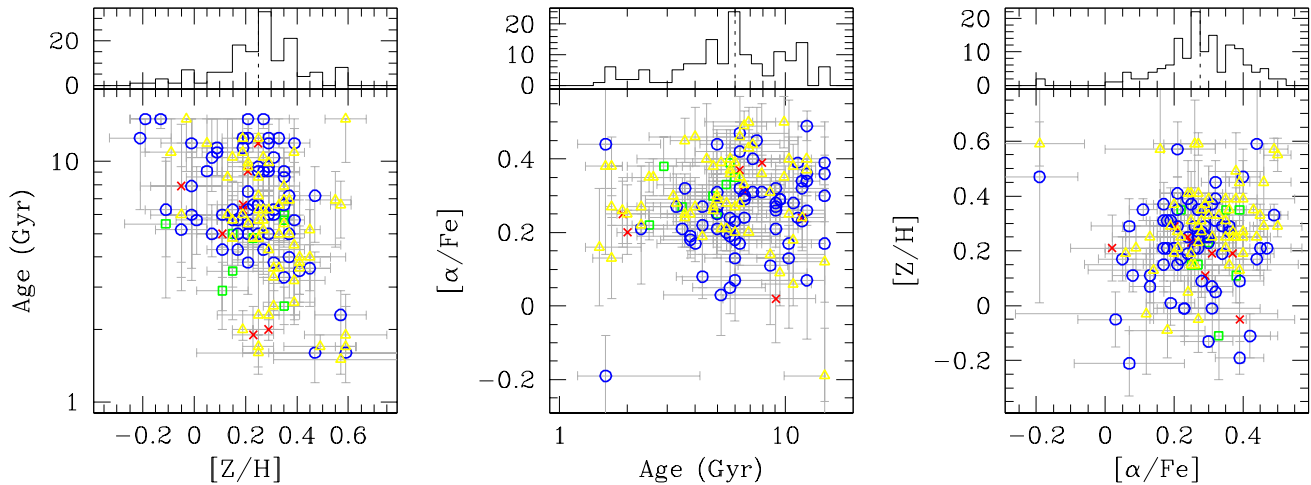}
\caption{The distributions of the SPPs for all galaxies in the cluster
  sample. Marginal distributions are shown for each parameter and the
  dotted lines show the median values of each parameter. Galaxies from
  Coma are shown as blue circles, from A1139 as green squares, from
  A3558 as yellow triangles and from A930 as red crosses.}
\label{gal params all}
\end{figure*}

The distributions of SPPs from the four clusters combined are shown in
Figure \ref{gal params all}. Marginal distributions are plotted for
each of the parameters and the median values are marked as dotted
lines. Galaxies from Coma are shown as blue circles, from A1139 as
green squares, from A3558 as yellow triangles and from A930 as red
crosses. Errors are shown for individual galaxies but two points must
be kept in mind. Firstly, the age and [Z/H] errors are correlated and
so the error bars here do not accurately reflect the true shape of the
confidence contours. For the two other combinations of parameters the
errors are much less correlated and so the error bars are a good
representation of the confidence contours (see Paper I). Secondly, a
galaxy's error estimate cannot be larger than its distance to the edge
of the model grid. This is only an issue for age estimates since we
quote two-sided errors. So, if a galaxy's error bar reaches 15 Gyr it
should be considered a lower limit only.

Looking at the distributions as a whole, we note that very few
galaxies have [Z/H] less than solar with most having $0.1\la
\mathrm{[Z/H]}\la 0.4$ dex. Similarly, $\afe$ is mostly greater than
solar and lies in the range $0.2\la \afe \la 0.4$ dex. The bulk of the
galaxies have old ages and almost all lie in the range $4\la
\mathrm{age} \la 15$ Gyr. The distributions we find for the cluster
sample are in general agreement with those found by previous authors
\citep[e.g.][]{gonzalez93, jorgensen99a, kuntschner00, trager00b,
  poggianti01, thomas05,collobert06}.

Looking at the SPP distributions of each cluster individually we find
few differences. The median values and errors for the SPPs in each
cluster are given in Table \ref{lowz median params}. The errors on the
median values were estimated from Monte Carlo bootstrap simulations in
the following manner. For each galaxy in a cluster, we randomly draw
index values from the galaxy's index error distributions, using the
same set of indices that were used to estimate the parameters
originally. These index values are then converted to parameter
estimates and the median values are determined. This process is
repeated 10,000 times and the RMS errors on the median values are
computed.

\begin{table}
\caption{The median SPP values for each cluster and the number of
  galaxies for which the parameters were measured.}
\begin{center}
\begin{tabular}{lcccr@{}l}
\hline\hline
Cluster & [Z/H] & Age & $\afe$ & \multicolumn{2}{c}{$N_{\mathrm{gal}}$}\\
\hline
Coma  & $0.25\pm 0.02$ & $6.6\pm 0.6$ & $0.27\pm 0.01$ & \hspace{1ex}6 & 5\\
A1139 & $0.19\pm 0.08$ & $4.9\pm 1.6$ & $0.32\pm 0.04$ &   & 8\\
A3558 & $0.29\pm 0.04$ & $5.7\pm 0.9$ & $0.29\pm 0.02$ & 6 & 1\\
A930  & $0.20\pm 0.08$ & $6.4\pm 2.1$ & $0.27\pm 0.05$ &   & 8\\
\hline
\end{tabular}
\label{lowz median params}
\end{center}
\end{table}

The distributions of SPPs from cluster-to-cluster show remarkable
similarity, with no significant differences between the median values
of any of the four clusters. However, to conclusively detect any
differences in the age and [Z/H] distributions, a two-sample 2D KS
test is required, due to the correlated errors between age and
[Z/H]. For the $\afe$ distributions a two-sample 1D KS test is
sufficient. The results of these tests are that the joint
distributions of age and [Z/H], and the distributions of $\afe$ in
each of the four clusters are consistent with each other.

Looking at the combined distributions of the cluster sample (142
galaxies in total), we find that the median [Z/H] is $0.25\pm 0.14$
dex, the median age is $6.0^{+4.6}_{-2.6}$ Gyr, and the median $\afe$
is $0.28\pm 0.11$ dex. The errors on these median values were
determined the same way as for the clusters individually. These median
values are shown as dotted lines in the marginal distributions in
Figure \ref{gal params all}.

There are some galaxies that have line-strengths that fall outside the
ranges predicted by the stellar population models, in that there are
six galaxies with age=15 Gyr, four with [Z/H]=0.59 dex, and two with
$\afe$=--0.19 dex. Two galaxies have more than one parameter on the
edge of the models and so in total there are 10 such galaxies. These
galaxies are assigned the combination of SPPs that most nearly fits
the combination of line-strengths, as mentioned in Paper I. It is hard
to determine the origin of this difference between the data and the
models, which could be due to errors in the stellar population models
(it should be noted that the models provide no estimate of systematic
uncertainties), or due to errors in the observations or data
reduction; only a small percentage of galaxies are affected, with
$\sim 90\%$ of galaxies having self-consistent model parameter fits.

For the galaxies with age=15 Gyr, it is possible that some of these
galaxies are affected by nebular $\hbeta$ emission, despite the fact
that care was taken in eliminating such galaxies. As was noted in
Paper I, it is possible for galaxies to have $\hbeta$ in emission and
no detectable [OIII]$\lambda$5007\AA\ \citep{nelan05}. If this
$\hbeta$ emission is undetectable, because it is swamped by $\hbeta$
absorption, and there is no [OIII]$\lambda$5007\AA\ emission, then
such a galaxy will not be classified as an emission-line galaxy and
will not be excluded from the cluster sample.

\begin{figure*}
\includegraphics[width=1.0\textwidth]{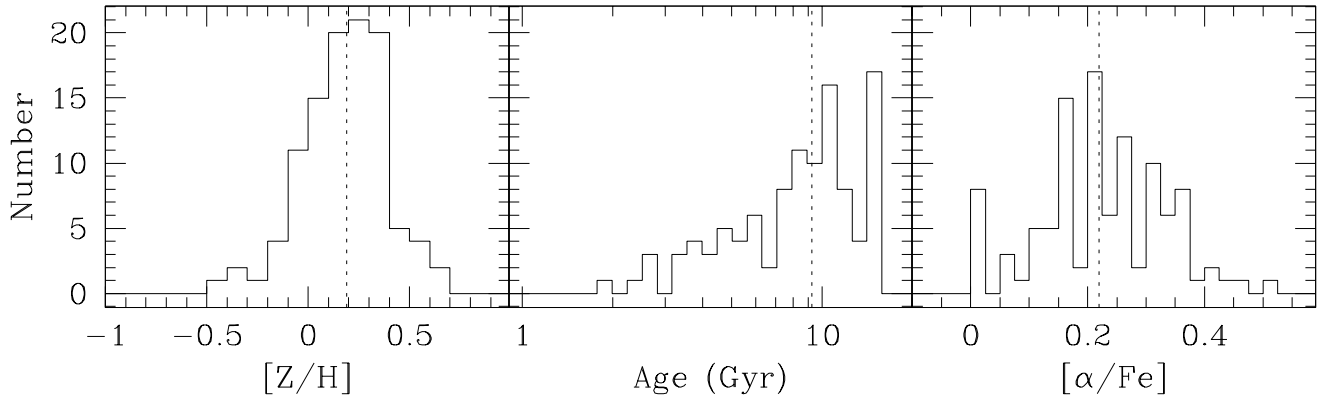}
\caption{The distributions of the SPPs, estimated from $\hbeta$,
  \mgb\ and Fe5335 only, for all galaxies in the cluster sample. The
  dotted lines show the median values of each parameter.}
\label{gal params all res}
\end{figure*}

It is also evident that there appears to be an anti-correlation
between [Z/H] and age, in the sense that younger galaxies are more
metal-rich. This might in principle be due to the non-orthogonal
nature of the [Z/H]-age grids produced by the stellar population
models, which mean that the errors on these two quantities are
correlated. The effect is that an increase in a [Z/H]-sensitive index
results not only in a increase in [Z/H] but also in a decrease in age;
correspondingly, an increase in an age-sensitive index results in a
decrease in age and an increase in [Z/H]. Due to the irregular shape
of the grid, this effect varies with location over the grid.

That a degree of anti-correlation between age and [Z/H] is introduced
by the correlated errors is not disputed; however, whether this
accounts for the anti-correlation entirely is still contentious
\citep{colless99a, jorgensen99a, trager00a, poggianti01, kuntschner01,
  terlevich02, proctor02, bernardi05, sanchez06}. Some studies find
that there exists a moderate anti-correlation over and above that
introduced by the errors \citep[e.g.][]{colless99a} while others find
that no correlation exists once the errors are taken into account
\citep[e.g.][]{kuntschner01}. To determine whether the
anti-correlation between age and [Z/H] is real or not, we perform a
simple test which involves comparing the degree of anti-correlation in
the observed joint age--[Z/H] distribution with those obtained by
Monte Carlo simulations.

Due to the need for an automated process to convert SPPs to
line-strengths (and vice versa) it is not feasible to use the method
of SPP estimation that makes use of all available indices. Therefore,
for this exercise, we rely on only three indices $\hbeta$, \mgb\ and
Fe5335. Combinations of these three indices are commonly used in
line-diagnostic diagrams to estimate SSPs. With this combination of
indices, the age distribution is approximately exponential and both
the [Z/H] and $\afe$ distributions are approximately Gaussian (see
Figure \ref{gal params all res}).

We start by generating a sample of galaxies with age, [Z/H], and
$\afe$ estimates (parameter triples); the [Z/H] and $\afe$ were drawn
from the distributions defined by the medians of the observed values
and their intrinsic scatters and the age was drawn from an exponential
distribution (see Section \ref{scatter} for details on how the
intrinsic scatters and the $e$-folding of the exponential age
distribution were determined). These parameter triples are then
converted to $\hbeta$, \mgb, and Fe5335 index values (index triples),
which are then perturbed using the observed indices error
distributions. These perturbed index triples are then converted back
to parameter triples and the Spearman rank correlation statistic $r_\mathrm{S}$
is calculated for age and [Z/H]. This procedure is carried out 10,000
times. From these Monte Carlo simulations we determine the probability
of obtaining by chance a correlation greater than that found for the
observed data, and we find that there exists a real anti-correlation
between age and [Z/H] (over and above the error-induced correlation)
that is significant at the $>3\sigma$ level.

\subsection{The distributions in the cluster-outskirts}
\label{low den dist}

\begin{figure*}
\includegraphics[width=1.0\textwidth]{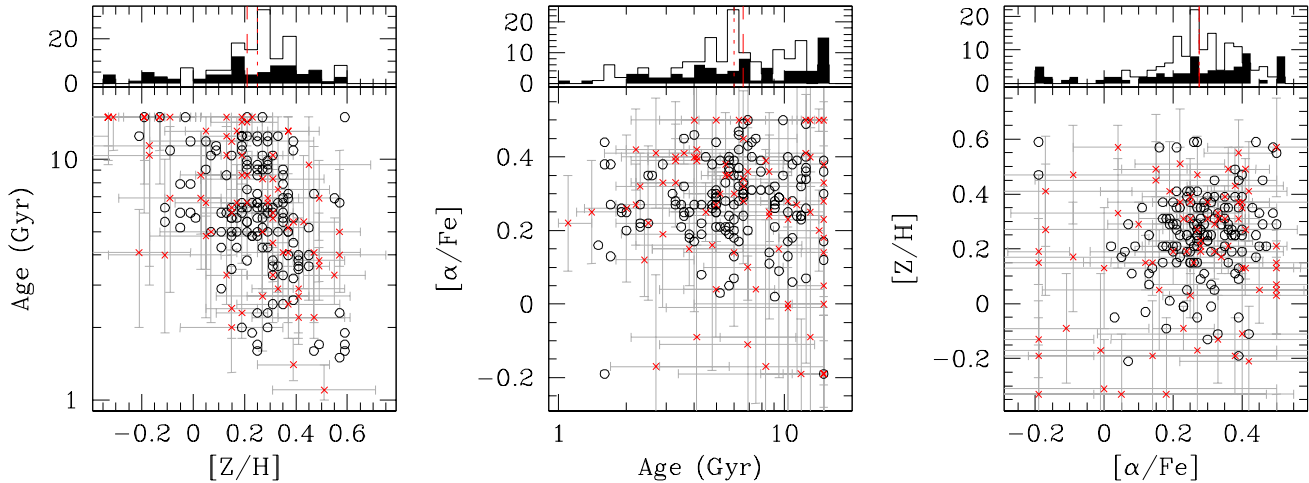}
\caption{Comparison of the SPPs in the cluster-outskirts sample (black
  circles) to those in the cluster sample (red crosses). The marginal
  distributions for the cluster sample are shown as open histograms
  and for the cluster-outskirts sample as black histograms. The dashed
  line in the marginal distributions represents the median parameter
  values in the cluster-outskirts sample and the dotted line that of
  the cluster sample.}
\label{gal params env}
\end{figure*}

The distributions of SPPs in the cluster-outskirts sample (red
crosses) are compared to those from the clusters (black circles) in
Figure \ref{gal params env}. For the sake of clarity, only the errors
for the cluster-outskirts sample are shown. The dashed lines in the
marginal distributions are the median galaxy parameters for the
cluster-outskirts sample and the dotted lines are for the cluster
sample. The galaxies in the cluster-outskirts sample have a median
[Z/H] of $0.21\pm 0.27$ dex, a median age of $6.6^{+8.1}_{-3.6}$ Gyr,
and median $\afe$ of $0.27\pm 0.21$ dex. Compared to the combined
cluster sample we detect no difference in the median values for the
galaxy parameters, although the errors are substantial.

Looking at the distributions we see that there is very little
difference between the galaxies in this sample and those in the
clusters, an assessment that is confirmed by KS testing (a two-sample
2D KS test in the case of the age and [Z/H] distributions and a
two-sample 1D KS test for the $\afe$ distributions).

The fact that the $\afe$ distributions are the same in the clusters
and their outskirts is intriguing, given that the cores of galaxy
clusters are thought to result from regions of high over-density where
the process of star formation occurs rapidly \citep{kauffmann98,
  schindler05, romeo05, delucia06}. Our result is, however, in
agreement with previous studies on the $\afe$ in low- and high-density
environments \citep{kuntschner02, thomas05}. These results suggest
that, in all environments, elliptical galaxies form on similar
timescales.

The often-found differences in low- and high-density environments
between the age \citep{trager00a, poggianti01, kuntschner02,
  terlevich02, caldwell03, proctor04a, thomas05} and [Z/H]
\citep{kuntschner02, thomas05}, with galaxies on average being $\sim
2$ Gyr younger and more metal-rich in lower-density environments, are
not reproduced here. Although, given the size of our errors it is not
surprising that we do not detect such a small difference. These
results are usually found comparing truly isolated galaxies with
cluster galaxies. As the galaxies in the cluster-outskirts sample are
actually drawn from the outer regions of the clusters and the
structures surrounding them (out to a projected radial distance of
$\sim 19$ $h_{70}^{-1}$ Mpc or $\sim10 R_\mathrm{vir}$), it is
possible that the contrast between the average densities of each
environment is not sufficient to reveal any differences in the stellar
populations, and that the change occurs at a lower density threshold
(e.g.\ between isolated galaxies and those in groups/filaments).

\begin{figure}
\includegraphics[width=0.5\textwidth]{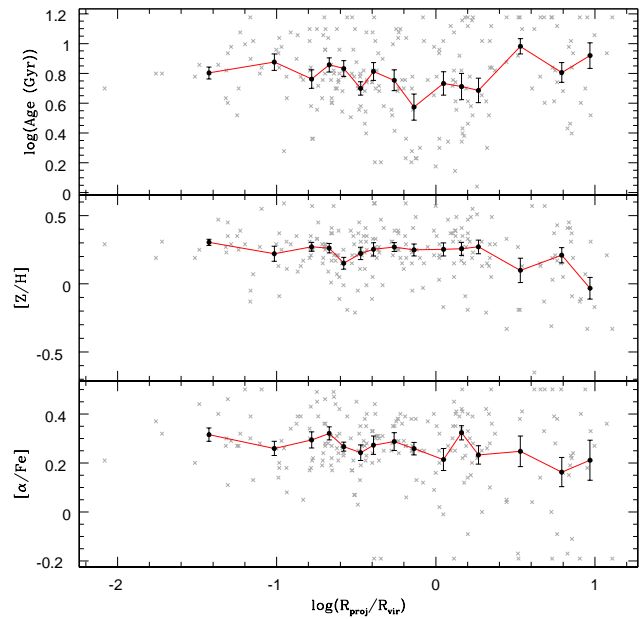}
\caption{The trend of mean age (top), mean [Z/H] (middle), and mean
  $\afe$ (bottom) as a function of cluster virial radius. The mean is
  calculated in bins containing 15 galaxies.}
\label{gal params r}
\end{figure}

Figure \ref{gal params r} shows the variations of the SPPs with
projected cluster-centric distance, normalised to the cluster virial
radius ($R_\mathrm{vir}$). The lines in these plots show the mean
values in bins containing 15 galaxies extending out to almost $10
R_\mathrm{vir}$. The error bars show the standard error of the mean
within each bin. We find no evidence of any significant trends,
although it does appear as if $\afe$ and age increase moving towards
the centre of the cluster, implying that the galaxies in the cluster
cores are older and have shorter star-formation timescales than those
in the cluster outskirts. Despite the lack of clear trends there are
two other interesting aspects of these plots. Firstly, there appears
to be a decrease in [Z/H] outside $\sim2 R_\mathrm{vir}$. Secondly, at
a distance slightly less than the cluster virial radius there appears
to be a dip in the average age. Such a dip would be expected in a
scenario where interaction with the ICM triggers star formation in an
in-falling galaxy, resulting in a stellar population characterised by
a young age and extended star-formation timescale. This decrease in
age begins at a radius $>3 R_\mathrm{vir}$ suggesting that the
influence of the cluster extends to large distances. We also note that
all galaxies with ages $\la 2$ Gyr are located near $R_\mathrm{vir}$
and the age of the youngest stellar population at a given radius
increases moving towards the cluster core.

\section{\boldmath The Parameter--$\sigma$ Relations}\label{scale}

\subsection{\boldmath The cluster relations}
\label{low clus rels}

\begin{figure*}
\begin{center}
\includegraphics[width=1.0\textwidth]{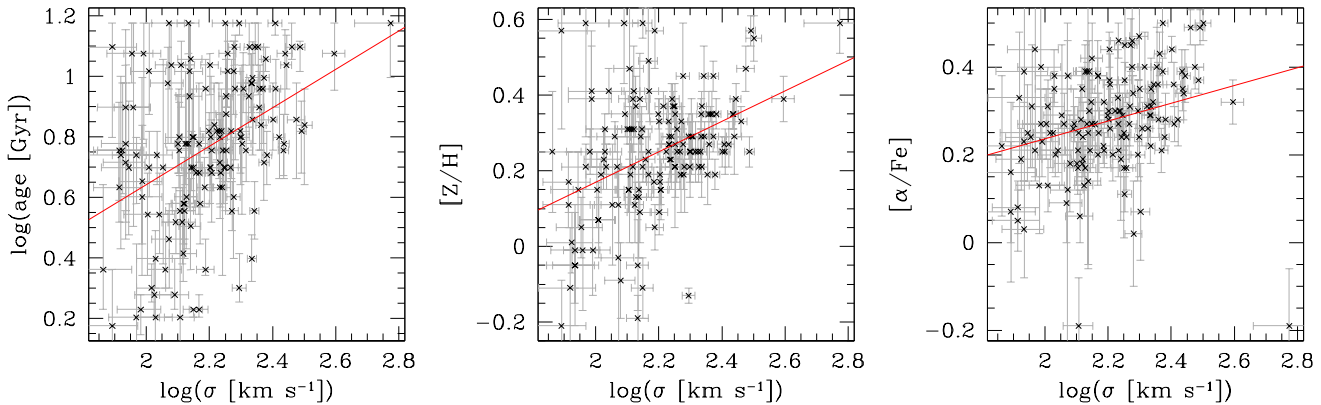}
\caption{The variations of the SPPs $\log(\mathrm{age})$ (left panel),
  [Z/H] (middle panel) and $\afe$ (right panel) with $\log\sigma$ of
  the galaxies in the cluster sample. All three parameters are
  moderately correlated with $\log\sigma$. The red line in each panel
  is the linear fit to the data.}
\label{param sig}
\end{center}
\end{figure*}

The variations of the SPPs with velocity dispersion are shown in Figure
\ref{param sig}. All parameters are found to be moderately correlated
with a high level of significance; $>5\sigma$ for $\log\mathrm{age}$
and $>4\sigma$ for [Z/H] and $\afe$. Performing linear fits to the
relations we find
\begin{eqnarray}
\log\mathrm{age} & = &
(0.64\pm0.12)\,\log\,\sigma-(0.63\pm0.26) \hspace{1ex},\label{age sig}\\
\mathrm{[Z/H]} & = &
(0.40\pm0.08)\,\log\,\sigma-(0.63\pm0.17) \hspace{1ex},\label{met sig}\\
\mathrm{[\alpha/Fe]} & = &
(0.20\pm0.06)\,\log\,\sigma-(0.17\pm0.13) \hspace{1ex}.\label{alpha sig}
\end{eqnarray}
These fits are shown as red lines in Figure \ref{param sig}. While
these results confirm the existence of the [Z/H]--$\sigma$ and
$\afe$--$\sigma$ relations, the correlation of age with velocity
dispersion found here is not so well accepted (see Section \ref{intro}
for references).

The behaviour of age and $\afe$ with velocity dispersion is
reminiscent of down-sizing \citep{cowie96}, where the typical mass of
a star-forming galaxy increases with redshift. Looking at the
distribution of ages, we see that at all velocity dispersions there
exist galaxies with very old stellar populations, but the age of the
youngest galaxy at a given velocity dispersion increases with velocity
dispersion. The peak epoch of star formation in more massive galaxies
thus occurred at higher redshifts.

\begin{table*}
\caption[Comparison of the SPP--$\sigma$
relations]{Comparison of the SPP--$\sigma$
relations from this study and the literature.}
\begin{center}
\begin{tabular}{lr@{.}lr@{.}lr@{.}lr@{.}lr@{.}lr@{.}l}
\hline
\hline
Reference            &  \multicolumn{4}{c}{$\log\mathrm{age}$}     & \multicolumn{4}{c}{[Z/H]}       & \multicolumn{4}{c}{$\afe$}\\
                     &  \multicolumn{2}{c}{Zero-point}             &  \multicolumn{2}{c}{Slope}                  &  \multicolumn{2}{c}{Zero-point}
                     &  \multicolumn{2}{c}{Slope}                  &  \multicolumn{2}{c}{Zero-point}             &  \multicolumn{2}{c}{Slope}\\
\hline
\multicolumn{13}{l}{This study}\\           
\hspace{5ex} Cluster & $-0$ & 63$\pm$0.26 & $0$    & 64$\pm$0.12 & $-0$    & 63$\pm$0.17 & $0$ & 40$\pm$0.08 & $-0$ & 17$\pm$0.13    & 0    & 20$\pm$0.06\\
\hspace{5ex} Cluster-outskirts & $-0$ & 43$\pm$0.45 & $0$    & 54$\pm$0.20 & $-0$    & 24$\pm$0.40 & $0$ & 18$\pm$0.18 & $-0$ & 52$\pm$0.31    & 0    & 33$\pm$0.14\\
\citet{thomas05} & $0$ & 46 & 0    & 24 & $-1$    & 06 & $0$ & 55 & $-0$ & 42 & 0    & 28\\
\citet{nelan05}      & \multicolumn{2}{c}{} & 0    & 59$\pm$0.13 & \multicolumn{2}{c}{} & $0$ & 53$\pm$0.08 & \multicolumn{2}{c}{} & 0    & 31$\pm$0.06\\
\citet{bernardi06}    & $-1$    & 72$\pm$0.31 & 1    & 15 & $-0$    & 64$\pm$0.01 & $0$ & 38 & $-0$ & 54$\pm$0.01    & $0$ & 32\\
\citet{smith06}    & \multicolumn{2}{c}{} & 0    & 72$\pm$0.14 & \multicolumn{2}{c}{} & $0$ & 37$\pm$0.08 & \multicolumn{2}{c}{}  & 0    & 35$\pm$0.07\\

\citet{graves07}     & \multicolumn{2}{c}{}                        & 0
& 35$\pm$0.03 & \multicolumn{2}{c}{} & 0 & 79$\pm$0.05 &
\multicolumn{2}{c}{} & 0    & 36$\pm$0.04\\
\citet{smith07}   & \multicolumn{2}{c}{} & $0$ & 64$\pm$0.12 &
\multicolumn{2}{c}{} & 0    & 38$\pm$0.09 & \multicolumn{2}{c}{} & 0
& 36$\pm$0.07 \\
\citet{thomas10} & $-0$ & 11$\pm$0.05 & $0$    & 47$\pm$0.02 & $-1$    & 34$\pm$0.04 & $0$ & 65$\pm$0.02 & $-0$ & 55$\pm$0.02    & 0    & 33$\pm$0.01\\
\hline
\end{tabular}
\label{par sig}
\end{center}
\end{table*}

Assuming that the scatter about the relations is Gaussian we find that
the intrinsic scatter in $\log\mathrm{age}$ is $0.20 \pm 0.02$ dex, in
[Z/H] is $0.10 \pm 0.02$ dex and in $\afe$ is $0.07 \pm 0.01$. The
uncertainty in these scatters were determined via Monte Carlo
simulations. Considering the mean observational errors are 0.14, 0.08,
and 0.08 dex, respectively, the tightness of these relations is
remarkable, and shows that the velocity dispersion is an excellent
indicator of the SPPs in early-type galaxies, with more massive
galaxies being older, more metal-rich and having shorter
star-formation timescales than less massive galaxies.

Table \ref{par sig} lists the SPP--$\sigma$
relations found in this work and in the literature. The literature
estimates come from a sample of early-type galaxies in high-density
environments \citep{thomas05}, a sample of red-sequence galaxies in
low-$z$ clusters \citep{nelan05}, a sample of early-type galaxies in
high-density environments from the SDSS \citep{bernardi06}, a sample
of red-sequence galaxies in low-$z$ clusters \citep{smith06}, a sample
of red-sequence galaxies drawn from the SDSS \citep{graves07}, a
sample of galaxies from the Shapley supercluster with $\sigma>100$ km
s$^{-1}$ \citep{smith07}, and a sample of red sequence galaxies in all
environments.

Overall, the slopes found here agree well with those found in the
literature. The slope of our $\log\mathrm{age}$--$\sigma$ relation is
consistent with most estimates in the literature except for that of
\citet{thomas05}, which is only a third as steep as ours, and that of
\citet{bernardi06} who find a slope twice as steep. The existence of a
correlation between age and velocity dispersion is still being debated
and this is reflected in the large variation in slopes found in the
literature. A possible cause for this is the difficulty in accurately
determining ages with the current set of models. The slope of our
[Z/H]--$\sigma$ relation agrees well with those found in the
literature, with exceptions of \citet{graves07} and \citet{thomas10}
who find a slope twice and 50 percent as steep, respectively. The
slope of our $\afe$--$\sigma$ relation is consistent with all of the
literature values despite being the smallest.

\subsection{The relations in the cluster-outskirts sample}\label{clus
  out rels}

The median values of the SPPs were found to be the same in both the
cluster sample and the cluster-outskirts sample, as were the SPP
distributions. In order to make a true comparison, one that is free
from any biases introduced by the two velocity dispersion
distributions, we now compare the parameter--$\sigma$ relations found
in the cluster sample with those in the cluster-outskirts sample.

\begin{figure*}
\includegraphics[width=1.0\textwidth]{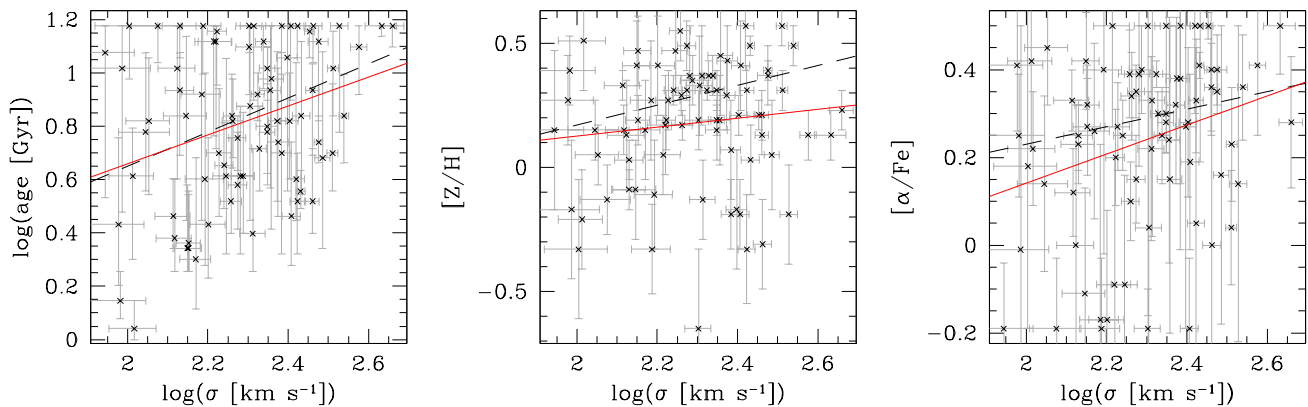}
\caption{The parameter--$\sigma$ relations for the cluster-outskirts
  sample. The solid red lines are the linear fits to the
  cluster-outskirts sample and the dashed black lines are the linear
  fits to the cluster sample.}
\label{param sig low}
\end{figure*}

Figure \ref{param sig low} shows the parameter--$\sigma$ relations for
the galaxies in the cluster-outskirts sample (solid red lines)
compared with those in the cluster sample (dashed black
lines). Details of the fits are given in Table \ref{par sig}.

\begin{figure*}
\includegraphics[width=1.0\textwidth]{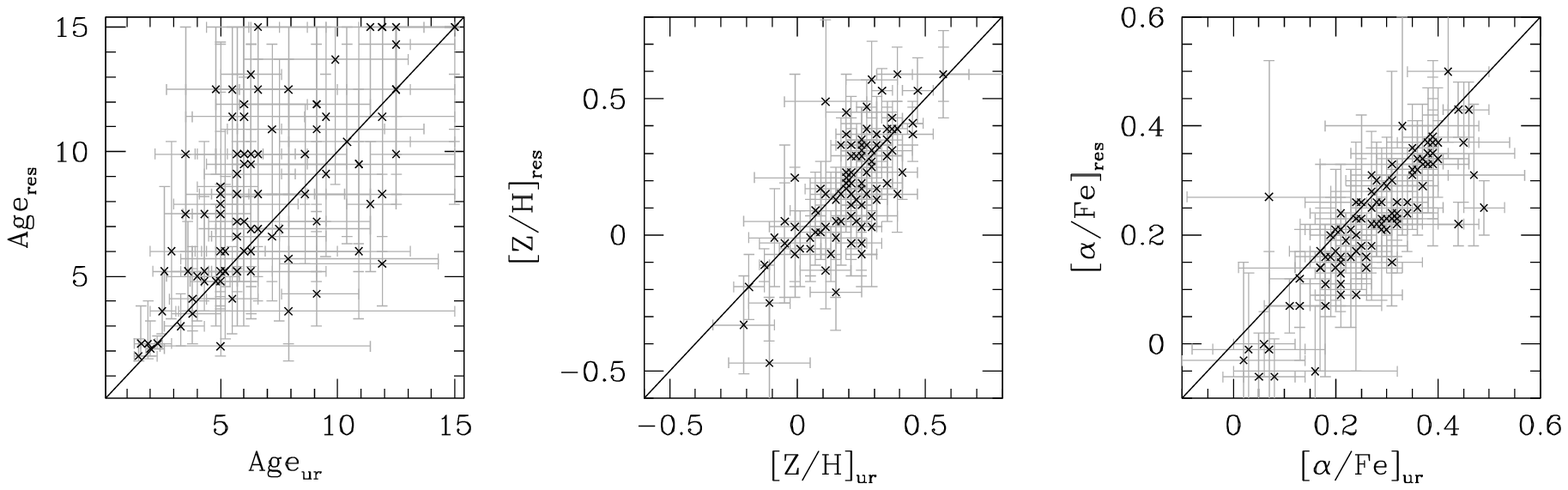}
\caption{Comparison of the age (left), [Z/H] (middle), and
  [$\alpha$/Fe] (right) estimates obtained from using the restricted
  and unrestricted set of indices. The solid lines represent the
  1-to-1 relation (see the text for details).}
\label{percent}
\end{figure*}

None of the SPPs are found to be significantly correlated with
velocity dispersion in the cluster-outskirts sample. Note that the
parameter errors are larger in the cluster-outskirts sample than the
cluster sample. This is due to the fact that the galaxies in the
cluster outskirts were observed with 6dF, which has less wavelength
coverage than 2dF, and therefore their SPPs were estimated, on
average, from fewer indices. It is possible that these larger errors
blur any correlation that may exist.

It is also possible that the change in the number and combination of
indices used to estimate the SPPs in the cluster-outskirts sample,
when compared to the cluster sample, is the reason why no correlations
are found in the former. We therefore check this effect by
re-estimating the SPPs of the cluster sample using only $\hbeta$,
\mgb\ and Fe5335. Using these estimates we find that all three
parameters are still correlated with velocity dispersion and that the
slopes of both relations are consistent within the errors. Therefore
it is unlikely that the lack of correlations in the cluster-outskirts
sample is a result of the change in the number of indices used. We
would like to point out that a fit (using the unrestricted set of
indices) was only accepted if $\hbeta$ was used and at least two other
indices. See Paper I for more details.

We show the results of comparing the two sets of estimates in Figure
\ref{percent}. For [Z/H] and $\afe$ the agreement between the two sets
of estimates is quite good, although there is an offset of 0.05 dex in
$\afe$ in the sense that the unrestricted set of indices gives larger
values of $\afe$. There is also an offset of 1.3 Gyr between the two
age estimates (but in the opposite sense to that of the $\afe$ offset)
and there is some scatter that increases with age. This is due to the
fact that, in index space, the distance between lines of constant age
decreases with increasing age so that small shifts in index strengths
can lead to large changes in age estimates. According to a Spearman
rank correlation test the two datasets are consistent. This implies
that the two ways of measuring the stellar population parameters rank
the objects in the same order and that there should be no effect on
our results.

\section{The Scatter in the Parameter Distributions in Clusters}
\label{scatter}

In part, the spread in the SPP distributions found for the cluster
sample is due to a combination of our observational errors and the
correlations with velocity dispersion. To determine the amount of
intrinsic scatter in the distributions, over and above that caused by
the observational errors, we ran a series of numerical
simulations. The method used was as follows.

Due to the need to create a large number of models containing a large
number of galaxies, which need to have their SPPs converted to
line-strengths (and vice versa), it is not feasible to estimate SPPs
using all available indices. Therefore, for this exercise, we rely on
the three indices used earlier in investigating the anti-correlation
between age and [Z/H], i.e. $\hbeta$, \mgb\ and Fe5335.

With this combination of indices, galaxies in the cluster sample have
an age distribution that is approximately exponential and [Z/H] and
$\afe$ have distributions that are approximately Gaussian (see Figure
\ref{gal params all res}). Therefore, we began by selecting a range of
$e$-foldings for the exponential age distribution ($\tau=0.10$, 0.15,
0.30, 0.55, 0.90, 1.60, 2.80, 5.00, 8.50, 15.00 Gyr), and ranges of
Gaussian scatters in [Z/H] ($\sigma_\mathrm{[Z/H]}=0.03$, 0.05, 0.08,
0.12, 0.20, 0.30, 0.50, 0.80, 1.25, 2.00 dex), and $\afe$
($\sigma_\mathrm{[\alpha/Fe]}=0.03$, 0.04, 0.07, 0.10, 0.15, 0.20,
0.30, 0.45, 0.70, 1.00 dex). Each range was chosen to be approximately
evenly spaced logarithmically so as to have better resolution at small
values. The medians of the cluster [Z/H] and $\afe$ distributions were
used as the means of their Gaussian distributions, and the exponential
age distribution was truncated at 5 Gyr (because the models are less
reliable at young ages) and at 14 Gyr (the age of the universe).

A model was generated for each combination of $\tau$,
$\sigma_\mathrm{[Z/H]}$, and $\sigma_\mathrm{[\alpha/Fe]}$ consisting
of 10,000 mock galaxies with ages, [Z/H], and $\afe$ (parameter
triples) drawn randomly from the specified distributions. If any of
the mock galaxies had a [Z/H] or $\afe$ that fell outside the extent
of the models\footnote{The \citet{thomas03a} models have $1\le t \le
  15$ Gyr, $-2.25\le \mathrm{[Z/H]}\le 0.67$ dex, and $0.0 \le
  \mathrm{[\alpha/Fe]}\le 0.5$ dex.} then they were assigned the
closest value from the models; this was not necessary for the ages
since they were truncated at the values stated above.

These parameter triples were converted, via the
\citeauthor{thomas03a}\ models, to the corresponding $\hbeta$, \mgb,
and Fe5335 index values (index triples). This was achieved as
follows. The models were divided up into cells defined by two adjacent
values of each SPP, and for each mock galaxy the cell in which it fell
was determined. Each cell is also defined by eight values of $\hbeta$,
\mgb, and Fe5335, corresponding to their values at the eight vertices
of the cubical cell. The index triple corresponding to the parameter
triple is then calculated by taking the mean of each of these eight
index values weighted by the distance in parameter space of the mock
galaxy from each corner of the cell.

These index triples were then perturbed by randomly drawing errors
from an observed galaxy's index error distributions. In doing this we
cycled through the observed galaxies so that each observed galaxy had
an approximately equal number of mock galaxies perturbed by its index
error distributions. The perturbed index triples were converted back
to parameter triples.

To ascertain how well the models fit the data a likelihood statistic
was used. The SPP space was divided up into bins and the probability
of each bin containing a galaxy was calculated. The age bins were 0.5
Gyr in width and ranged from 0 and 15 Gyr, the [Z/H] bins were 0.125
dex in width and ranged from $-2.2875$ to 0.7125 dex, and the $\afe$
bins were 0.025 dex in width and ranged from 0.0 to 0.5 dex. This
resulted in 30 age bins, 24 [Z/H] bins, and 20 $\afe$ bins. Before the
likelihood statistic was calculated, the model was lightly smoothed to
minimise the number of bins with zero probabilities; a 3D Gaussian,
with $\sigma=2/3$ bins for all dimensions, was used as the smoothing
kernel, and bins within 3$\sigma$, in each dimension, were used in
calculating the value of the bin probability in the smoothed model
(i.e.\ two bins either side in each dimension were used). The
likelihood statistic is obtained by determining which bins the
observed galaxies lie in and summing the logarithm of the bin
probabilities.

Monte Carlo simulations were used to estimate the probability of
obtaining a likelihood statistic larger than the observed one by
chance. There are 103 galaxies with measurements of the indices
required to estimate the SPPs in the cluster sample, so we simulate
103 galaxies in an identical manner to the model and calculate the
maximum likelihood statistic. This is done 10,000 times. The fraction
of simulations with likelihood statistics larger than the observed
value are recorded and 3D confidence contours are generated.

\begin{figure*}
\includegraphics[width=0.49\textwidth]{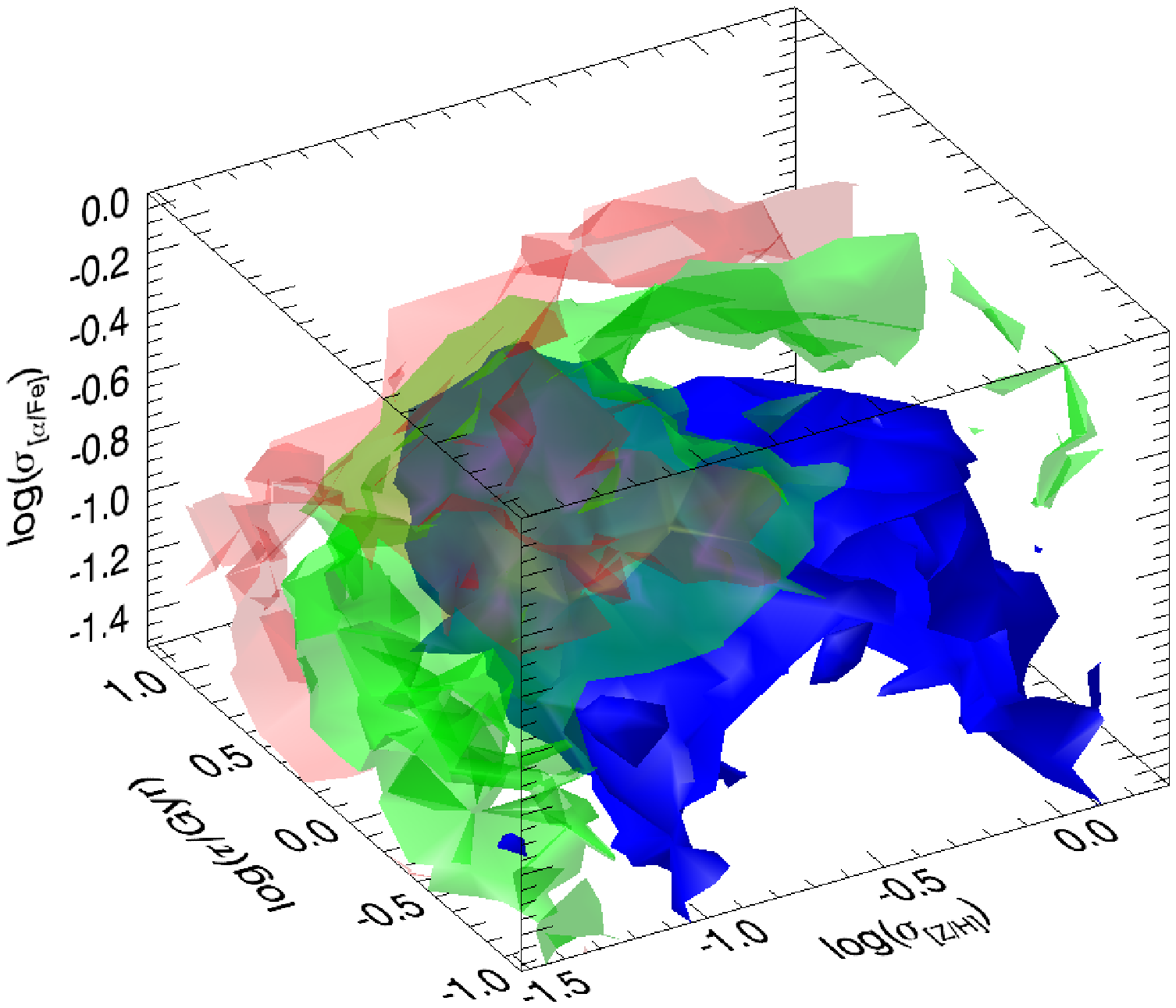}
\includegraphics[width=0.49\textwidth]{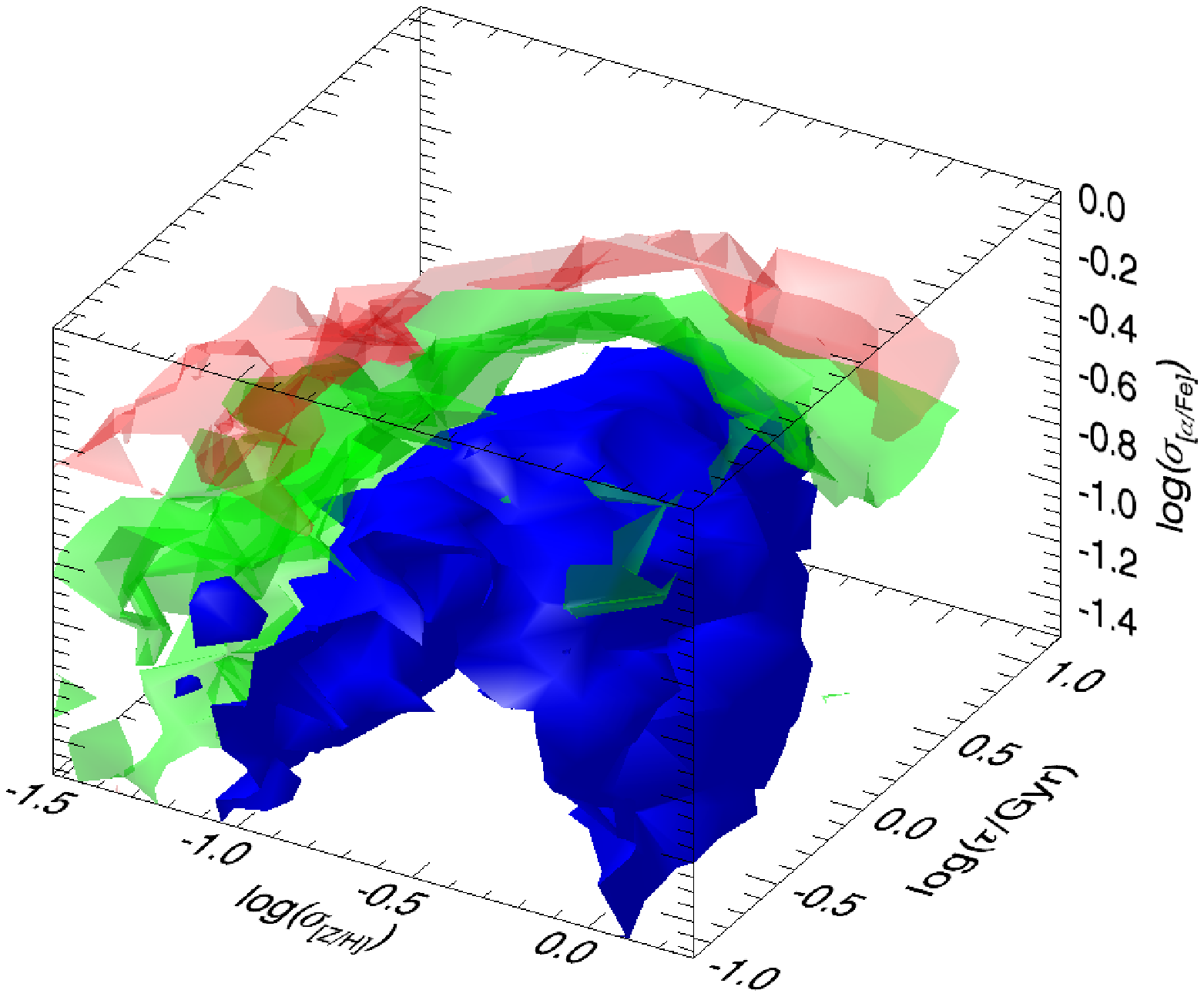}
\caption{Two views of the 3D confidence contours of the $e$-folding of
  the exponential distribution of ages ($\tau$) and the RMS scatters
  in the Gaussian distributions of [Z/H] and [$\alpha$/Fe]
  ($\sigma_\mathrm{[Z/H]}$ and $\sigma_\mathrm{[\alpha/Fe]}$),
  generated from the numerical simulations described in the text. The
  blue, green, and red surfaces are the 1$\sigma$, 2$\sigma$, and
  3$\sigma$ confidence contours.}
\label{3d contour}
\end{figure*}

\begin{figure*}
\includegraphics[width=1.0\textwidth]{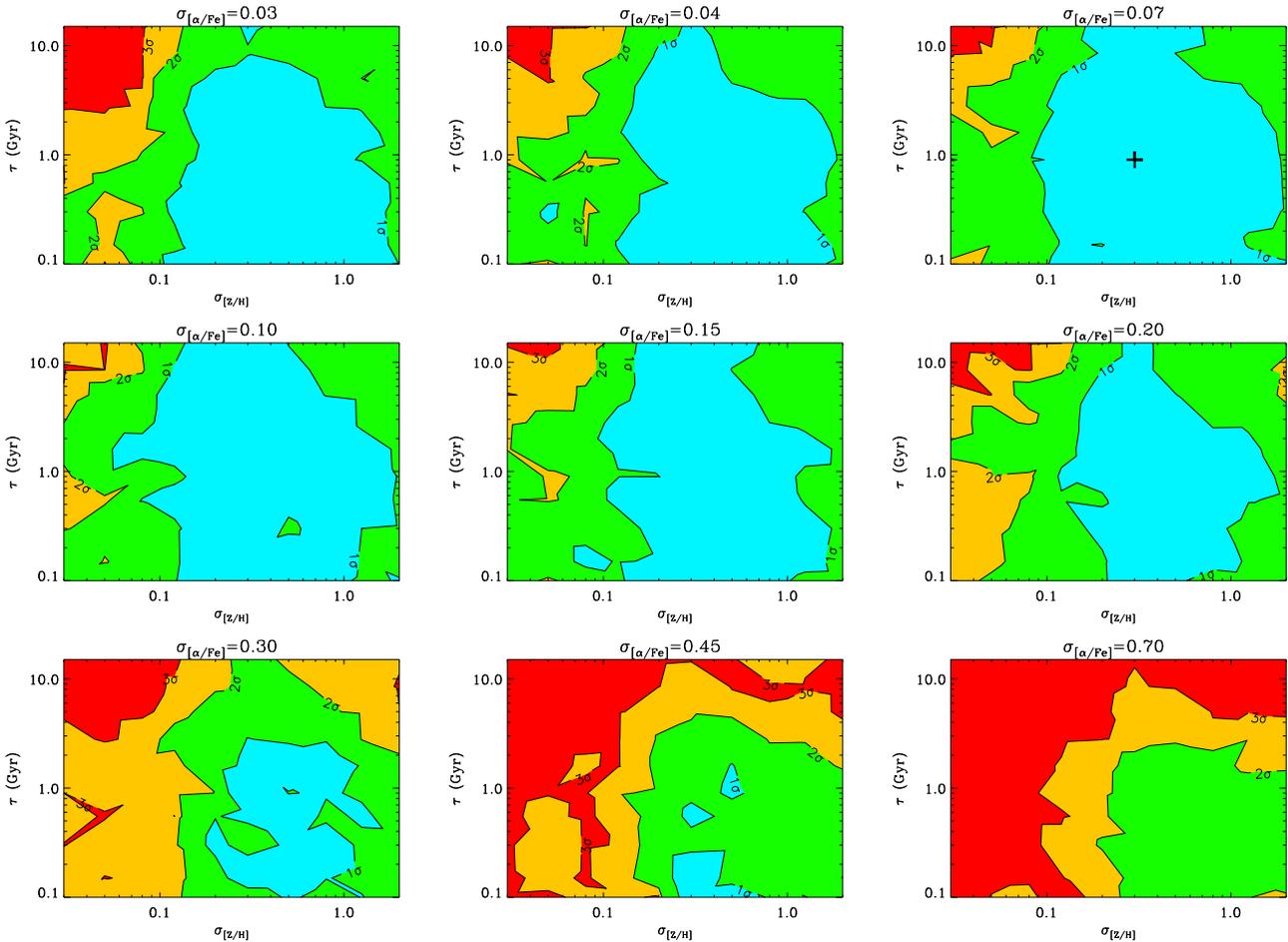}
\caption{Slices, parallel to the age-[Z/H] plane, through the 3D
  confidence contours shown in Figure \ref{3d contour}. Blue is
  $<1\sigma$, green 1--2$\sigma$, orange 2--3$\sigma$, and red
  $>3\sigma$. The value of the [$\alpha$/Fe] scatter used to generate
  the model is shown at the top of each panel. The cross in the panel
  with $\mathrm{[\alpha/Fe]}=0.07$ marks the model that is most
  consistent with our data.}
\label{contour slices}
\end{figure*}

These 3D confidence contours are shown in Figure \ref{3d contour},
which presents the likelihood of our dataset being represented by a
model with the specified $e$-folding of the exponential distribution
of ages and specified RMS scatters in the Gaussian distributions of
[Z/H] and $\afe$. While Figure \ref{3d contour} provides an overall
picture of the acceptable scatters in the SPPs, more detailed
information can be revealed by taking slices through the 3D confidence
contours. Figure \ref{contour slices} shows slices taken parallel to
the age-[Z/H] plane at values of $\afe$ corresponding to those used in
generating the models.

These simulations provide us with a great deal of information
regarding the intrinsic scatter in each of the SPPs. Firstly, it is
difficult to constrain the $e$-folding of the exponential age
distribution. This is due to small uncertainties in line-strengths
translating to large changes in the age estimate combined with
relatively large observational errors in the $\hbeta$ index. Although
models with small values of $e$-folding are not ruled out, we find
that our data is most consistent with the model having $\tau = 900$
Myr. Secondly, small scatters in [Z/H] ($\sigma_\mathrm{[Z/H]}<0.1$
dex) are strongly ruled out, as are large scatters
($\sigma_\mathrm{[Z/H]}>2.0$ dex). We find the model with
$\sigma_\mathrm{[Z/H]} = 0.3$ dex is the most consistent with our
data. Finally, large scatters in $\afe$ ($\sigma_{\afe}>0.3$ dex) are
also strongly ruled out. Models with low scatters in $\afe$ are not
ruled out, but we find that the most consistent model is that with
$\sigma_{\afe} = 0.07$ dex. The model that gives the maximum
likelihood (i.e.\ $\tau\sim 900$ Myr, $\sigma_\mathrm{[Z/H]}\sim 0.3$
dex, and $\sigma_{\afe} = 0.07$ dex) is marked by a cross in Figure
\ref{contour slices}.

Since the SPPs were found to be correlated with velocity dispersion,
and since the galaxies in the cluster sample have a range of velocity
dispersions, it is understandable that we should detect an intrinsic
scatter in the SPP distributions. The degree to which the intrinsic
scatter in the parameter distributions is attributable to trends with
velocity dispersion can be estimated by comparing the scatter expected
given the velocity dispersion distribution in the cluster and the
intrinsic scatter about the parameter--$\sigma$ relations.

We use the parameter--$\sigma$ relations to convert the galaxies'
velocity dispersions into parameter values. The scatter in these
values is then calculated; the intrinsic scatter about the relation
(as found in Section \ref{low clus rels}) is then added in quadrature
to the scatter due to the relation with velocity dispersion, and the
result is compared to that obtained from the simulations described
above. We do this only for [Z/H] and $\afe$, which were found to have
approximately Gaussian distributions.

The scatter expected in [Z/H] on this basis is 0.21 dex which is
comparable to the intrinsic scatter of $\sigma_\mathrm{[Z/H]}\sim 0.3$
dex estimated above. The expected scatter in $\afe$ is 0.06 dex which
is very close to the estimated intrinsic scatter $\sigma_{\afe} \sim
0.07$ dex. It appears then that the intrinsic scatter in both the
[Z/H] and $\afe$ distributions can almost entirely be accounted for by
the parameter--$\sigma$ relation and the intrinsic scatter about it.

For [Z/H], the scatter due to the correlation is 0.16 dex while the
intrinsic scatter about the relation is 0.13 dex; thus, most of the
intrinsic scatter in the [Z/H] distribution comes from the correlation
with velocity dispersion. For $\afe$, the scatter due to the
correlation is 0.03 dex while the intrinsic scatter about the relation
is 0.05 dex; thus most of the intrinsic scatter in the $\afe$
distribution comes from the intrinsic scatter about the
$\afe$--$\sigma$ relation.

\section{Discussion}\label{discuss}

The process by which galaxies form and the factors with the greatest
influence on their evolution are unresolved issues. In models of
galaxy formation based on hierarchical merging in a $\Lambda$CDM
universe \citep[e.g.][]{delucia06} the merger history of an early-type
galaxy can fall anywhere between two extremes: mergers could occur
early-on and rapidly with subsequent passive evolution---the revised
monolithic collapse scenario \citep[e.g.][]{merlin06}; or the galaxy
could experience a more prolonged history of mergers---the extended
merging scenario \citep[e.g.][]{toomre77}. The environment of a galaxy
and its mass are the two most influential factors on galaxy evolution
but their effects need to be to disentangled to gain an insight into
the relative importance of each.

For cluster galaxies, the numerical simulations performed here show
that the intrinsic scatters in [Z/H] and $\afe$, and the $e$-folding
of the exponential distribution of ages, are all quite small, as would
be expected from a rapid formation at high redshift followed by
passive evolution. Additionally, positive correlations are found
between all three of the SPPs and velocity dispersion. The
correlations found for both [Z/H] and $\afe$ confirm well-known trends
\citep{trager00a, kuntschner01, proctor02, thomas02, caldwell03,
mehlert03, nelan05, thomas05, bernardi06, gallazzi06, kelson06,
sanchez06, thomas10}. However the correlation found for age is not
widely recognised; some authors find age and velocity dispersion to be
positively correlated \citep{proctor02, proctor04a, proctor04b,
nelan05, thomas05, gallazzi06, bernardi06, thomas10} while others find
they are not correlated at all \citep{trager00a, kuntschner01,
terlevich02}. Not only do we find that age is positively correlated,
but also that it is the most significantly correlated out of the three
SPPs, being significant at the $5\sigma$ level. These correlations are
qualitatively consistent with more recent semi-analytic models
\citep{delucia06}. These models incorporate feedback from AGN
\citep{croton06}, which heats the available gas, preventing further
star formation in massive galaxies and bringing the predicted scaling
relations into agreement with those observed.

There is a level of homogeneity to early-type cluster galaxies
evidenced in the similarity of their line-strength and SPP
distributions and the consistency of the correlations between these
quantities and velocity dispersion that is found between the clusters
studied here. That this homogeneity is maintained amongst clusters of
varying richness and morphology indicates that differences in the
cluster environment have relatively little effect on the stellar
populations of early-type galaxies. The dominant factor appears to be
mass.

While mass plays a large role in determining the SPPs in early-type
galaxies, it does not do so completely. The size of the role played
varies; while [Z/H] is almost entirely a function of velocity
dispersion, $\afe$ is less so. The intrinsic scatters in [Z/H] and
$\afe$ in the cluster galaxies were found to be almost entirely
accounted for by the scatter produced by the correlations with
velocity dispersion and the intrinsic scatter about these
relations.

Galaxies in low-density environments are observed to be, on average,
$\sim 2$ Gyr younger than galaxies in clusters \citep{trager00a,
poggianti01, kuntschner02, terlevich02, caldwell03, proctor04b,
denicolo05, thomas05, bernardi06, sanchez06}. The situation is not so
clear with regards to differences in [Z/H] and $\afe$.  Recent models
predict that galaxies in denser environments are older and more
metal-rich than isolated galaxies \citep{delucia06}. We do not find
significant differences in the line-strength distributions or the SPP
distributions between the cluster-outskirts sample and the cluster
sample. However, the size of our SPP errors make it difficult to
detect the small differences in age reported by other authors and any
difference may be masked by the 1.3 Gyr offset to older ages found for
galaxies that their parameters estimated from fewer indices; such
galaxies are found predominantly in the cluster-outskirts sample. We
do find that the tight correlations with velocity dispersion of both
the line-strengths and SPPs that are found in clusters are weaker in
the cluster outskirts, suggesting that the modes of formation in the
cluster outskirts are more varied than those in the cluster cores.

Similar results were found by \citet{thomas10} who conclude that the
formation and evolution of early-type galaxies are relatively
insensitive to environment and are instead driven by self-regulation
processes and their intrinsic properties such as mass. However, our
result that the SPP trends with velocity dispersion are weaker in the
cluster-outskirts than in the clusters is in conflict with these
results \citep[see also][]{bernardi98,bernardi06}. The dip in ages
found near the cluster virial radius (see Figure \ref{gal params r})
suggests that movement into the cluster environment may induce a burst
of star formation and, combined with the down-sizing evident in Figure
\ref{param sig}, that this burst may occur preferentially in lower
mass galaxies (similar to the rejuvenated population from
\citeauthor{thomas10}). Presumably, our two samples (cluster and
cluster-outskirts) are combined in their single high-density sample,
which may result in the masking of this trend. Our results are broadly
consistent with the results of \citet{smith06} who find a variation of
the SPPs with cluster radius. \citeauthor{thomas10} find that
environment becomes more of an influence moving to less massive
galaxies \citep[see also][]{haines06}.

Our results can be best explained by the revised monolithic collapse
model, but is also consistent with the extended merging model if the
mergers are dissipationless \citep[see][]{delucia06}. \citet{pipino09}
have a model of elliptical galaxy formation that implements a detailed
treatment of chemical evolution, and they find that the correlation
between $\afe$ and velocity dispersion is much less steep than that
observed. \citeauthor{pipino09} find that AGN quenching can help to
improve the agreement, but this worsens their mass-metallicity relation
compared to observations. They suggest that both relations can be
reproduced provided the formation of all spheroids happens
quasi-monolithically, i.e. the formation of the stars occurs in
sub-units and that this star formation and the assembly of the
sub-units into an elliptical galaxy occur at the same time and in the
same place.

The star-formation rates in galaxies are found to increase with
increasing distance from the cluster centre, and to converge to the
mean field rate at distances greater than $\sim 3 R_\mathrm{vir}$
times the virial radius \citep[][see also
\citet{gomez03}]{lewis02}. We find evidence that there is a dip in the
mean ages of galaxies just inside the cluster virial radius, possibly
due to secondary star formation that reduces the luminosity-weighted
mean age. This could be evidence that in-falling galaxies, upon
reaching the virial radius of a cluster, undergo a burst of star
formation triggered by the dense intra-cluster medium. If such a burst
only amounted to a small fraction of the galaxy's total mass then its
effect on the integrated light would be short-lived and the galaxy
would rapidly return to appearing old and red. Similar to the findings
of \citeauthor{lewis02}, we find that this decrease in age actually
begins at $>3 R_\mathrm{vir}$, suggesting that the influence of the
cluster environment extends to large distances. Whereas the result of
\citeauthor{lewis02} was based mainly on late-type galaxies, the
remarkable thing about this result is that it applies to early-type
galaxies.

There are two sources of bias that we must check have not influenced
the above results: the number of indices used to derive the SPPs and
and the aperture corrections. It is possible that the lack of
correlations with velocity dispersion for the SPPs in the
cluster-outskirts sample is due to the fact that, on average, fewer
indices were used to estimate them than those in the cluster
sample. This results in larger SPP errors and might potentially
explain the lack of correlations with velocity dispersion. However, as
we show in Section \ref{clus out rels}, using only three indices to
estimate the SPPs does not significantly alter their distributions
compared to those derived from an unrestricted set of indices.

It is also possible that mass segregation within the cluster could
cause the weakening of the line-strengths in the cluster outskirts, if
that is where less massive galaxies are preferentially found. This is
a result of the same aperture correction being applied to a galaxy
irrespective of how massive it is and despite the fact that the
line-strengths being corrected were measured within differing
effective radii. We find that our two samples (cluster and
cluster-outskirts) have a similar shape to their magnitude
distributions but that the galaxies in the cluster-outskirts are on
average $\sim 1$\,mag brighter, due to the facts that there are more
galaxies in the cluster sample and that we target the brightest
galaxies. So the correlation of weak line-strengths for less massive
galaxies cannot be the reason for the weakening of the trends with
velocity dispersion in the cluster-outskirts galaxies. We see no
reason, therefore, not to believe that the weakening of the relations
in the cluster-outskirts is real.

The overall picture that emerges from this study is as
follows. Early-type galaxies in clusters form a homogeneous class of
objects that form in a process similar to a revised monolithic
collapse and whose stellar populations are largely determined by their
velocity dispersion (mass) and are relatively unaffected (at least
differentially) by the cluster environment. The more massive they are
the older and more metal-rich they are, and the shorter their
star-formation timescales. The stellar populations of early-type
galaxies in the outskirts of clusters, in contrast, appear less
influenced by mass because, due to the varying environments they
formed and evolved in, their evolutionary histories are more varied
and this causes the correlations with mass to be less significant. A
galaxy, especially a less massive one, that is falling into a cluster
will, upon nearing the virial radius, undergo a burst of star
formation. Once these stars have faded and ceased dominating the
integrated light the galaxy will appear indistinguishable from those
in the cluster core.

\section{Summary}\label{summary}

In summary, we have measured velocity dispersions, redshifts and
absorption-line strengths for a magnitude-limited
($\mathrm{b}_{\mathrm{J}} \lid 19.45$) sample of galaxies drawn from
four clusters (Coma, A1139, A3558, and A930 at $\left< z
\right>=0.04$) and their surrounds (extending to $\sim10
R_\mathrm{vir}$). Using the fully-calibrated absorption-line indices
coupled with the stellar population models of \citet{thomas03a}, we
have estimated ages, [Z/H], and $\afe$ for 219 galaxies. We have used
these data to investigate the effects of mass and environment on the
stellar populations of early-type galaxies and our results can be
summarised as follows.

\begin{enumerate}

\item For galaxies in the cluster sample, all indices are positively
  correlated with velocity dispersion, with the exceptions of $\hbeta$
  and $\hbetag$, which are negatively correlated, and Fe5406, which is
  uncorrelated.
\item Only \mgb\ and $\hbetag$ are correlated with velocity dispersion
  in the cluster-outskirts sample. The slopes of these two relations
  are consistent with those found for the cluster sample.
\item The cluster cores are relatively free from young galaxies and
  from galaxies that have experienced recent star formation. These
  galaxies are more commonly found outside $R_\mathrm{Abell}$.
\item The stellar populations in clusters form a homogeneous
  population. Despite the fact that our sample was drawn from four
  clusters spanning the ranges of Bautz-Morgan classifications and
  Abell richness classes, the line-strength--$\sigma$ relations,
  parameter-$\sigma$ relations, and SPP distributions are consistent
  between clusters.
\item There is no difference between the line-strength distributions
  and SPP distributions in the clusters and their outskirts.
\item The SPPs in clusters are correlated with velocity dispersion,
  suggesting that more massive galaxies are older, have shorter
  star-formation timescales, and are more metal-rich. These
  correlations are found to be weaker in the cluster outskirts.
\item For galaxies in the cluster sample, the $e$-folding time of the
  age distribution is 900 Myr, the intrinsic scatter in the [Z/H]
  distribution is 0.3 dex, and the intrinsic scatter in the $\afe$
  distribution is 0.07 dex. These latter two intrinsic scatters can
  almost entirely be accounted for by the correlations with velocity
  dispersion and the scatter about the relations. We conclude,
  therefore, that the mass of a galaxy plays a major role in
  determining its stellar populations.

\end{enumerate}

Further high quality observations of galaxies at higher redshifts will
allow the development of a consistent model of early-type galaxy
formation at all masses and in all environments.\\

The authors thank the anonymous referee for a careful reading of the
paper and for the many helpful suggestions that improved it
considerably.

\bsp

\label{lastpage}

\end{document}